\documentclass[twocolumn,aps,prb,showpacs,eqsecnum,floatfix]{revtex4}
\usepackage{graphicx,bm}
\begin{document}
\title{Bogoliubov - de Gennes versus Quasiclassical description of
Josephson structures}
\date{15 December 2001 }
\author{M. Ozana}
\author{A. Shelankov$^{*}$}
\author{J. Tobiska}
\affiliation{Department of Theoretical Physics, Ume{\aa} University, 901
87 Ume{\aa}, Sweden}

\begin{abstract}
The applicability of the quasiclassical theory of superconductivity in
Josephson multi-layer structures is analyzed. The quasiclassical
approach is compared with the exact theory based on the Bogoliubov -
de Gennes equation. The angle and energy resolved (coarse-grain)
currents are calculated using both techniques. It is shown that the
two approaches agree in $SIS'IS''$ geometries after the coarse-grain
averaging.  A quantitative discrepancy, which exceeds the
quasiclassical accuracy, is observed when three or more interfaces are
present. The invalidity of the quasiclassical theory is attributed to
the presence of closed trajectories formed by sequential reflections
on the interfaces.  \end{abstract}

\pacs{74.80.Dm, 74.80.-g, 74.50.+r, 74.20.Fg}

\maketitle  

Studying charge current through weak links, the Josephson effect, is
one of the most important part of superconductivity, both theory and
experiment \cite{Lik79,Tinkham}. Besides general interest, this
problem is important for engineering the numerous devices based on the
Josephson effect.  The Josephson effect reveals itself in tunneling
junctions as well as more complex mesoscopic structures built of
superconducting and normal layers.  A major part of the theoretical
results in this field has been obtained using the method of
quasiclassical Green's functions
\cite{Eil68,LarOvc,Sch81,SerRai83,BelWilBru99}.  The advantage of this
general method is that disorder and inelastic processes can be
conveniently incorporated into the theory.  In a ballistic case, one
can apply a more simple technique \cite{She80b,BTK82} based on the
Bogoliubov - de Gennes (BdG) equation \cite{deG89}. A strong side of
the BdG-approach is that it is valid for description of the interface
reflection and transmission, where the potential varies on a
microscopic length and the quasiclassical theory in its original form
fails.  As shown by Zaitsev \cite{Zai84}, an isolated partially
transparent interface can be taken into account by a proper boundary
condition for the quasiclassical Green's functions.  Recently, there
has been considerable technical progress where the boundary condition
is formulated using the Schopohl-Maki parameterization \cite{SchMak95}
of the quasiclassical Green's function \cite{Esc00}, or in terms of
effective wave functions \cite{SheOza00,LucEckShe01}.  However, it has
been argued \cite{OzaShe01,OzaShe02} that these boundary conditions
may give a wrong result in the case when a coherent scattering by
several interfaces takes place.  It is shown in Ref. \cite{OzaShe01}
that the quasiclassical density of states of a $SS'$ sandwich
disagrees with the ``exact'' one, found from the Gor'kov equation.
The disagreement has been attributed in Ref.\cite{OzaShe02} to the
presence of closed trajectories formed by sequential reflections by
the interface and the outer boundaries of the sandwich. In particular,
the correction to the quasiclassical Green's function due to the
loop-like trajectories violates the normalization condition, which is
an essential element of the quasiclassical technique. In this paper,
we extend these results to the case of an open geometry and analyse
applicability of the quasiclassical theory to the Josephson effect in
a multi-layer mesoscopic structure.

The quasiclassical theory is a simplified version of the ``exact''
theory of superconductivity based on the Gor'kov Green's function
formalism. The main assumption made in the course of its derivation is
that the potentials vary slowly on the Fermi wave length
$\lambdabar_{F} = 2\pi / p_{F}$, $p_{F}$ being the Fermi momentum,
that is the parameter $\lambdabar_{F}/ \xi_{0}$ is small, where
$\xi_{0}$ is the coherence length ($\xi_{0} \sim v_{F}/ \Delta $,
$v_{F}$ is the Fermi velocity).  The question we address in this paper
is whether there are corrections to the theory which are not
controlled by the quasiclassical parameter $\lambdabar_{F}/ \xi_{0}\ll
1$.

\begin{figure}
\centerline{\includegraphics[width=0.9\columnwidth]{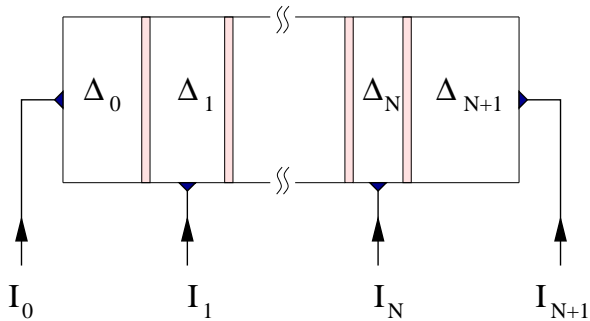}}
\caption{ 
A planar, multi-layer structure consisting of $N$ layers separated by
barriers and sandwiched by half-infinite electrodes. The complex order
parameter
 denoted  $\Delta_{0}, \ldots ,
\Delta_{N+1}$ are considered as inputs. 
It is assumed that each layer is connected to an independent current
source $I_{0}, I_{1}\ldots$ so that one achieves any given distribution
of the phase of the order parameters without violation of the current
conservation .
} 
\label{fig:multi-layer}
\end{figure}

To judge if the quasiclassical approach gives valid results, we
compare its predictions with the solution to the BdG equation.  In the
clean case of the mean-field theory, the BdG approach is fully
equivalent to the Green's function method, which is the starting point
to the derivation of the quasiclassical approximation. For this reason
we consider the BdG solutions as ``exact'' for the purpose of the
comparison.  More specifically, we consider a multi-layer
$SS'S''\ldots$ structure shown in Fig.\ref{fig:multi-layer} and
calculate the angle and energy resolved partial current $j(\theta ,
\varepsilon)$, where $\theta $ is the angle of incidence and
$\varepsilon $ is the energy of the excitation propagating through the
multi-layer structure.  Comparing the results of the two approaches,
we make our conclusions on the validity of the quasiclassical
approximation.

\begin{figure}
\centerline{\includegraphics[width=0.8\columnwidth]{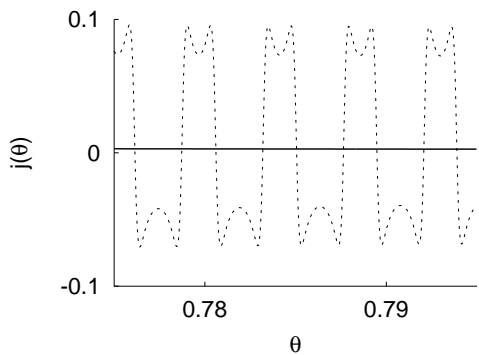}}
\caption{
Angle resolved current in a SISIS structure as a function of angle
$\theta$. The energy $\varepsilon =1.2 \Delta_{0}$. The order
parameter in the leftmost superconductor is $\Delta_{0}$. The order
parameters in the next layers read $\Delta_{1} = e^{i\pi/4}
\Delta_{0}, \Delta_{2} = i \Delta_{0}$.  The thickness of the internal
layer is $a = v_{F}/\Delta_{0}$.  The interface transparencies
$T_{1}=0.1, T_{2} = 0.5$.  The dashed and solid lines show the BdG
$j_{BdG}$ and the quasiclassical current $j_{qc}$, respectively.  }
\label{SSS} \end{figure}

As discussed in detail in Ref.\cite{OzaShe01,OzaShe02}, one cannot
make any conclusions comparing directly $j_{BdG}(\theta ,
\varepsilon)$, evaluated from the BdG-equation with its quasiclassical
counterpart $j_{qc}(\theta , \varepsilon)$. The point is that in the
BdG approach, the incident particle is taken as a plane wave with
precisely defined wave vector whereas in the quasiclassical approach
one deals with classical trajectories, where the momentum in the
direction perpendicular to the velocity has quantum uncertainty. The
infinitely extended plane wave suffers multiple reflections on the
interfaces, and the reflected/transmitted waves inevitably interfere
because of the infinite extension.  The interference leads to an
intricate picture of Fabry-Perot like resonances and a fine structure
in the angle dependence on the scale $\delta \theta \sim
\lambdabar_{F}/ a$ where $a$ is the layer width.  To illustrate this
point, we show in Fig. \ref{SSS} angle-energy resolved $j_{BdG}$
current through a SISIS-system of superconductors (S) separated by two
barriers (I) in a narrow region of angles (the BdG calculations are
done by the method presented below, see also \cite{BriGol00}). The
``exact'' current shows rapid and strong fluctuations in the region
where the quasiclassical current is almost a constant. However, on a
large scale of angles, $j_{BdG}$ averaged in a small angle window
(coarse-grain current) is a smooth function.  As discussed in
\cite{OzaShe01,OzaShe02}, the coarse-grain averaging is equivalent to
building stationary wave packets, peaked on classical trajectories, on
which the quasiclassical theory is formulated.  It is on this low
resolution level where the quasiclassical and exact theory are
expected to agree with each other.  For these reasons, we use only
coarse-grain BdG-current for comparison with the quasiclassical
theory. For definiteness, we calculate the current at the leftmost
interface of the multi-layer structure.

In principle, the coarse-grain averaging can be performed analytically
applying the path length expansion method of Ref.\cite{OzaShe02}. (In
case of a double-layer, the averaging can be done directly with the
result expressed via the elliptic integrals \cite{AshAoyHar89,Nag98}.)
Nevertheless, we perform the averaging by a numerical integration to
avoid lengthy algebra.

The paper is organized as follows.  In Sect.\ref{sec:BdG} we discuss
solutions to the Bogolubov - de Gennes equation. In Sect.\ref{plane},
we build plane waves and then, in Sect.\ref{Scat}, we introduce the
scattering S-matrix.  In Sect. \ref{jBdG}, we express the BdG current
via the elements of the S-matrix. The method which we use to evaluate
the current in the quasiclassical technique is presented in
Sect.\ref{jQC}.  Numerical results are shown in Sect. \ref{res}, and
their interpretation is presented in Sect.\ref{diss}.  In
Sect.\ref{conc} we discuss validity of our results in more realistic
models. Technical details of the derivation are collected in
Appendices.

\section{Bogolubov-de Gennes Equation} \label{sec:BdG}

In this section, we consider the theory of multi-layer structure in
the framework of the Bogoliubov - de Gennes equation.  Stationary
two-component wave function $\psi(\bm{r}) = {u \choose v}$ of an
excitation with the energy $E$ satisfies the BdG-equation
\cite{deG89}, $ \hat{H}\psi = E \psi $,
\begin{equation}
\hat{H}=
\left(
\begin{array}{lr}
\xi( \bm{p} - {e\over c}{\bm A} ) + V & \Delta \\
\Delta^{*} & - \xi (\bm{p} + {e\over c}{\bm A} ) - V
\end{array}
\right) 
\label{1}
\end{equation}
where $\xi (\bm{p})= \frac{\bm{p}^2}{2m}- \frac{p_{F}^2}{2m}$, $p_{F}$
being the Fermi momentum, $V(\bm{r})$ is the potential energy, $\Delta
(\bm{r}) $ is the complex order parameter, and $\bm{A}(\bm{r})$ is the
magnetic vector potential.

The charge current density $\bm{J}$ can be found as
\begin{equation}
\bm{J} = \Re 
\left(\psi^{\dagger} \hat{\bm{J}}\psi  \right) \; 
\label{bjc}
\end{equation}
where $\hat{\bm{J}}= - c\,\frac{\partial \hat{H}}{\partial \bm{A}}$ is
the current operator 
\begin{equation}
\hat{\bm{J}} = 
\frac{e}{m}\left(\bm{p} - 
\hat{\tau_{z}} {e\over c} \bm{A} \right)\;, 
\label{ajc}
\end{equation}
$\tau_{z} $ being  the Pauli matrix.

The non-diagonal elements of current operator $\hat{\bm{J}}(\bm{r})$, 
$\langle \psi_{n}|\hat{\bm{J}}(\bm{r})|\psi_{n'}\rangle $ are evaluated as
\begin{equation}
\langle \psi_{n}|\hat{\bm{J}}(\bm{r})|\psi_{n'}\rangle =
\frac{1}{2}
\left( \hat{\bm{J}}\psi_{n}\right)^{\dagger} \psi_{n'}
+ \frac{1}{2} \psi_{n}^{\dagger}\left( \hat{\bm{J}}\psi_{n'}\right)
\; ;
\label{smc}
\end{equation}
In superconductors, the charge current created by an elementary
excitation in a state $\psi_{n}$ is not a conserving quantity, {\it
i.e.} $\text{{\bf div}}\bm{J}_{nn}\neq 0$. The charge conservation is
restored after the summation over all the BdG excitations, provided
the pair potential $\Delta $ is self-consistent.

To take advantage of the unitarity property, one considers the
conserving {\em quasiparticle} current, $\bm{j}^{\text{qp}}(\bm{r})$,
calculated with the help of the operator, $\hat{\bm{j}}^{\text{qp}}=
{\partial H\over{\partial \bm{p}}}$, that is
\begin{equation}
\hat{\bm{j}}^{\text{qp}} = 
\frac{1}{m} \left( \hat{\tau_{z}}\bm{p}  - 
 {e\over c} \bm{A} \right) \;, 
\label{4ic}
\end{equation}
and 
\begin{equation}
\bm{j}^{\text{qp}}(\bm{r}) =
\Re \; \psi^{\dagger}\hat{\bm{j}}^{\text{qp}}\psi \;.
\label{7ic}
\end{equation}
The continuity equation
\begin{equation}
\text{{\bf div}}\, \bm{j}^{\text{qp}} =0  \;.
\label{8ic}
\end{equation}
follows from the BdG equation.

We solve the BdG equation for the case of a planar structure shown in
Fig.\ref{fig:multi-layer}.  It is composed of $N$ layers surrounded by
two half-infinite homogeneous superconductors. The complex order
parameters in each layer is a constant which is taken as an
independent input.

Choosing the $x$ axis perpendicular to the layer plane, the partially
transparent interfaces are modelled by the $\delta$-function barrier
\begin{equation}
V=  \frac{\lambda }{m} \delta(x)\;, 
\label{qmc}
\end{equation}
where $\lambda$ is the strength of the potential, $m$ is the
mass. Below, we characterize the interface by the parameters $R=
|\lambda / (p_{F}+ i \lambda) |^2$ and $T$ ($R+T=1$) which have
the meaning of the reflection coefficient ($R$), and transparency($T$)
for the normal incidence.

\subsection{Plane wave solutions}\label{plane}

Due to the in-plane translational invariance, the solutions can be
taken in the following form
\begin{equation}
\Psi(\bm{r}; \bm{p}_{||}) = e^{i \bm{p}_{||} \cdot \bm{r}} \psi(x) \; ,
\label{2}
\end{equation}
where $\bm{p}_{||}$ is the in-plane momentum, and $\psi $ obeys the
one-dimensional BdG equation.

\begin{figure}
\centerline{\includegraphics[width=0.8\columnwidth]{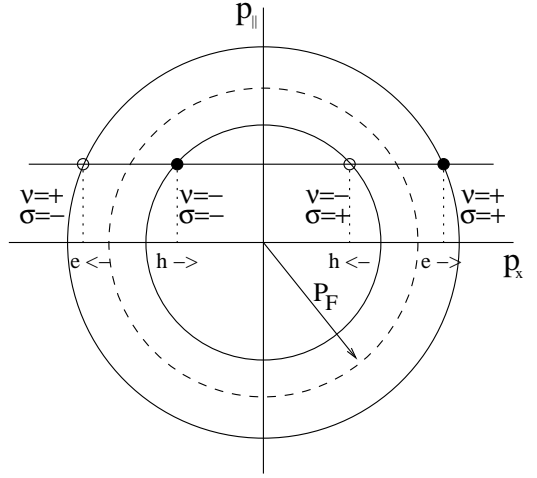}}
\caption{4 linearly independent solutions to BdG equation. Index $\nu$
describes the type of the quasiparticle $\nu=+$ is the electron-like
and $\nu=-$ the hole-like excitation; $\sigma$ defines the sign of the
$x-$component of the momentum.  As shown by arrows, the excitation
propagates to the right if $ \nu \cdot \sigma = +1$, and to the left
otherwise.
} \label{fig:momentsII} \end{figure}

First, we consider the plane wave solution to the BdG equation in each
of the layers where $\Delta=const$ and $V=0$.  The function $\psi(x)$
satisfies the equation
\begin{equation}
\left(
\begin{array}{cc}
\hat{\xi}_{x}& \Delta \\
\Delta^{*} & - \hat{\xi}_{x}
\end{array}
\right) \psi(x) = 
E
\psi(x) \; ,
\label{3}
\end{equation}
where $\hat{\xi}_{x} = \left(\hat{p}_{x}^2 + \bm{p}_{||}^2 -
p_{F}^2\right)/2m$, and $\hat{p}_{x} = -i\, d/dx$.  Eq.(\ref{3}) has 4
linearly independent plane wave solutions:
\begin{equation}
\psi_{\nu \sigma}(x) =
 e^{i\, \sigma p_{\nu} x}
\psi_{\nu} \; , \; \nu = \pm\; , \; \sigma=\pm \; 
\label{5}
\end{equation}
where the momentum $p_{\nu }$ is found from (see Fig. \ref{fig:momentsII})
\begin{equation}
p_{\nu} = \sqrt{\bm{p}_{F}^{2} - \bm{p}_{||}^{2} \pm 2 m  \xi} \; , \;
 \Re \; p_{\pm} > 0
\label{4}
\end{equation}
with
\begin{equation}
\xi =  \sqrt{ E^{2} - |\Delta|^{2}} \; .
\label{mjc}
\end{equation}
We choose the branch of the square root so that $\xi > 0$ at $E>
|\Delta |$ and $\Im \xi >0 $ when $E < |\Delta |$

The two-component amplitudes $\psi_{\pm}$ found from Eq.~(\ref{3})
with $\hat{\xi }_{x}$ substituted for $\xi $, may be chosen in the
following form
\begin{equation}
\psi_{+} =  \frac{1}{c  }\;{1 \choose  a  }
\quad ; \quad
\psi_{-} =  \frac{1}{c  }{b\choose  1 }
\label{njc}
\end{equation}
 where 
\begin{equation}
a  = \frac{\Delta^{*}}{E + \xi}\quad,\quad 
b = \frac{\Delta}{E + \xi}\quad,\quad
c = \sqrt{\frac{2 \xi}{E + \xi }}
\; ,
\label{yjc}
\end{equation}
($c=\sqrt{1- a b}$). These expressions are applicable for any energy
$E$ including the gap region $E< |\Delta |$.  Outside the gap, $b = a
^{*}$.

The amplitudes are normalized to the unit flux, {\it i.e.}
\begin{equation}
\psi_{\pm }^{\dagger}\hat{\tau_{z}} \psi_{\pm }= \pm 1 \quad,\quad E> |\Delta |
\; .
\label{pjc}
\end{equation}
Note the orthogonality relation (outside the gap),
$\psi_{\pm}^{\dagger} \hat{\tau_{z}} \psi_{\mp}= 0$, in agreement with
the current conservation in Eq.~(\ref{8ic}).  For any $E$, the
determinant of the matrix $[ \psi_{+}, \psi_{-}]$ the columns of which
are $\psi_{+}$ and $\psi_{-}$, equals to unity (in other words,
$\psi_{+}^{T}i \hat{\tau}_{y}\psi_{-}=1$).

For future needs, we define conjugated amplitudes
$\psi^{\ddagger}_{\nu }$,
\begin{equation}
\psi^{\ddagger}_{\nu }\equiv  - \nu\, \psi_{-\nu}^{T} \,i\hat{\tau_{y}} 
\; ,
\label{qjc}
\end{equation}
which posses  useful properties of orthonormality and
completeness:
\begin{equation}
\psi^{\ddagger}_{\nu } \psi_{\nu'}= \delta_{\nu \nu '}
\quad,\quad 
\sum\limits_{\nu}\psi_{\nu}\psi^{\ddagger}_{\nu } = \hat{1}
\label{rjc}
\end{equation}
where $\hat{1}$ is the unit $2\times 2$ matrix.  Outside the gap,
where $\psi_{\nu}^{\ddagger} = \nu\, \psi_{\nu}^{\dagger}
\,\hat{\tau_{z}}$, the orthonormality expresses the quasiparticle
current conservation.

The physical meaning of the quantum numbers $\nu $ and $\sigma $ is
clear: in accordance with the sign of the probability flux
Eq.~(\ref{pjc}), the excitations with $\nu =1$ are electron-like
whereas the states $\nu =-1$ are hole-like. The parameter $\sigma $
shows the direction of the momentum. The excitations for which the
product $\sigma \nu$ equals to +1 (-1) propagate in the positive
(negative) direction of the $x-$axis.

\subsection{The scattering matrix}\label{Scat}

To introduce the scattering matrix, we first define the in-coming and
out-going free states, the amplitudes of which are related by the
S-matrix.  We number the incoming plane wave states in the following
way:
\begin{equation}
\begin{array}{ccl}
|1 \rangle^{(0)}_{\text{in}} = \Psi_{++}^{(L)} &\quad,\quad &
|2 \rangle^{(0)}_{\text{in}} = \Psi_{--}^{(L)} \\
|3 \rangle^{(0)}_{\text{in}} = \Psi_{+ -}^{(R)} & \quad,\quad &
|4 \rangle^{(0)}_{\text{in}} = \Psi_{-+}^{(R)} \;.
\end{array}
\label{ckc}
\end{equation}
As before, the first of the lower indices of $\Psi$'s specifies the
electron-hole degree of freedom (``+'' for electron, and ``-'' for
hole) and the second one shows the direction of the momentum $\sigma =
\pm$; the upper index $L$ or $R$ specifies the initial location of the
excitation on the left or right side of the structure. The out-going
states are
\begin{equation}
\begin{array}{ccl}
|1 \rangle^{(0)}_{\text{out}} = \Psi_{+-}^{(L)} &\quad,\quad &
|2 \rangle^{(0)}_{\text{out}} = \Psi_{-+}^{(L)} \\
|3 \rangle^{(0)}_{\text{out}} = \Psi_{+ +}^{(R)} & \quad,\quad &
|4 \rangle^{(0)}_{\text{out}} = \Psi_{--}^{(R)} \;.
\end{array}
\label{dkc}
\end{equation}

The basis wave function read
\begin{equation}
\Psi_{\nu \sigma }= \sqrt{\frac{m}{2\pi p_{\nu}}}
e^{i \sigma p_{\nu}x + i \bm{p}_{||} \bm{\cdot r}}  \psi_{\nu } \;,
\label{rmc}
\end{equation}
with $\psi_{\nu }$ from Eq.~(\ref{njc}), these function are normalised
to $\delta(\bm{p}_{||}-\bm{p}_{||}') \delta (E - E')$.

In the L- or R-regions of free motion, the solution to the BdG
equation, $|i \rangle $, corresponding to the incident quasiparticle
in the state $|i \rangle^{(0)}_{\text{in}}$ can be presented as
\begin{equation}
|i \rangle = |i \rangle^{(0)}_{\text{in}} + \sum\limits_{f=1}^{4}
S_{f i}|k \rangle^{(0)}_{\text{out}}
\; .
\label{ekc}
\end{equation}
These equations with $i=1,\ldots,4$ define the $4 \times 4$ S-matrix.
The method which allows us to evaluate the elements of S-matrix is
presented in Appendix \ref{trans}.

\subsection{The BdG current}\label{jBdG}

Expressed via the distribution function of the excitations $n_{i}(
E)$, \big( where $E$ is the energy and $i=1,\ldots,4$ and $E$ is the
quantum number introduced in Eq.~(\ref{ekc})\big), the charge current
in the $x-$ direction reads \cite{deG89}
\begin{equation}
J(x) = \int d{\bm{p}_{||}}  \int\limits_{0}^{\infty } 
dE \sum\limits_{i=1}^{4}(2 n_{i}( E)-1) \langle i,E|\hat{J}(x)| i,E \rangle 
\; ,
\label{jkc}
\end{equation}
$\hat{J}(x)$ being the current operator Eq.~(\ref{ajc}).
We restrict ourself to the simplest case where the distribution
function depends only on energy, {\it i.e.} $n_{i}(E)= n(E)$. Then,
the current can be written as 
\begin{equation}
J(x) = \int d{\bm{p}_{||}}  \int\limits_{0}^{\infty } 
dE \left(2 n(E)-1 \right) J(E, \bm{p}_{||}; x)
\label{xkc}
\end{equation}
where the partial current density, $J(E, \bm{p}_{||}; x)$ at the point
$x$, is
\begin{equation}
J(E, \bm{p}_{||};x) =  
\sum\limits_{i=1}^{4}\langle i,E|\hat{J}(x)| i,E \rangle
\label{ykc}
\end{equation}

To evaluate the current in the left or right electrodes, we substitute
$|i \rangle $ from Eq.~(\ref{ekc}), and take into account the
unitarity property, $\sum\limits_{i=1}^{4} S_{f'i}^{*}S_{fi}=
\delta_{f'f}$. We get
\begin{equation}
J(E, \bm{p}_{||};x) = 
J_{E}^{(0)}(x)  + 2\Re\sum\limits_{i,f=1}^{4}S_{fi}^{*}J_{fi}^{(0)}(x)
\label{nkc}
\end{equation}
where $J_{E}^{(0)}(x)= \sum\limits_{k=1}^{4}J_{kk}(x)$: here summation
is performed over the 4 plane wave states Fig. \ref{fig:momentsII} on
the left or right side of the structure.  The meaning of $J_{E}^{(0)}$
is that it would give the (partial) current in the left or right
region if the plane wave states with the given energy $E$ were equally
occupied.  This is the contribution to the current which produces the
bulk supercurrent $2eN_{s} \bm{v}_{s}$.  In our case, $J_{E}=0$ since
the phase of the order parameter is assumed to be a constant within
the outside regions.

The second term in the right hand side of Eq.~(\ref{nkc}) is due to
the interference of the incoming and outgoing waves.  Considering for
definiteness the left region, the initial states $i=1,2$ interfere
with the final $f= 3,4$ states.  For the energy $E$ outside the gap,
$J_{fi}^{(0)}$ is other than zero only if $i=1, f=2$ or $i=2, f=1$
(i.e. for the interference with the Andreev reflected particle).  The
partial current density $J(E, \bm{p}_{||})$ at the point adjacent to
the first interface, {\it i.e.} at $x=0^{-}$ reads
\begin{equation}
J(E, \bm{p}_{||}) = 
   2\Re 
\left(
S_{21}^{*}J_{41}^{(0)}(0^{-})
+
   S_{12}^{*}J_{32}^{(0)}(0^{-})
 \right) \; .
\label{pkc}
\end{equation}
Calculating the current matrix elements Eq.~(\ref{tjc}) with the
amplitudes in Eqs.~(\ref{njc}), and (\ref{Mmnb}), one derives from
Eq.~(\ref{pkc}) that
\begin{equation}
J(E, \bm{p}_{||}) = 
   \frac{1}{\pi  } \Re \;\frac{1}{ \xi } 
\left(
S_{21}\, \Delta_{0}
-
   S_{12}\,\Delta_{0}^{*}
 \right) \quad,\quad  E > |\Delta_{0}| \; .
\label{2kc}
\end{equation}
where $\Delta_{0}$ is the order parameter in the left electrode (see
Fig. \ref{fig:multi-layer}) \cite{neglect}.  The scattering matrix is
calculated by the transfer matrix method as described in Section
\ref{trans}.

Eq.~(\ref{2kc}) gives the BdG current carried by plane wave states
with definite value of $\bm{p}_{||} $. This quantity strongly
fluctuates (see e.g. Fig. \ref{SSS}) as a function of the incidence
angle $\theta $, $p_{||}= p_{F}\sin \theta $ in the region $\Delta
\theta \sim 1/( p_{F} a)$, $a$ being typical interlayer distance.  To
come to the trajectory-like picture, one performs averaging in a
region of angles $\sim \Delta \theta $. It is the coarse-grain current
which should be compared with the current density found from the
quasiclassical technique.

\section{Quasiclassical Current}\label{jQC}

\begin{figure}
\centerline{
\includegraphics[width=0.42\columnwidth]{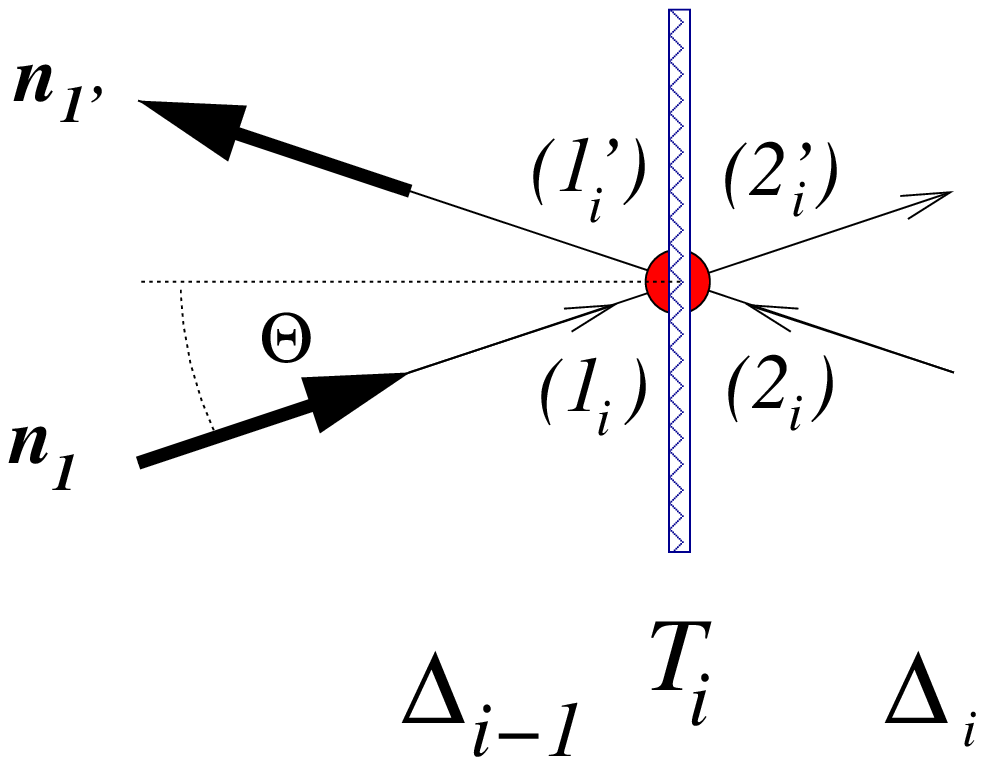}
\hspace{0.05\columnwidth}
\includegraphics[width=0.42\columnwidth]{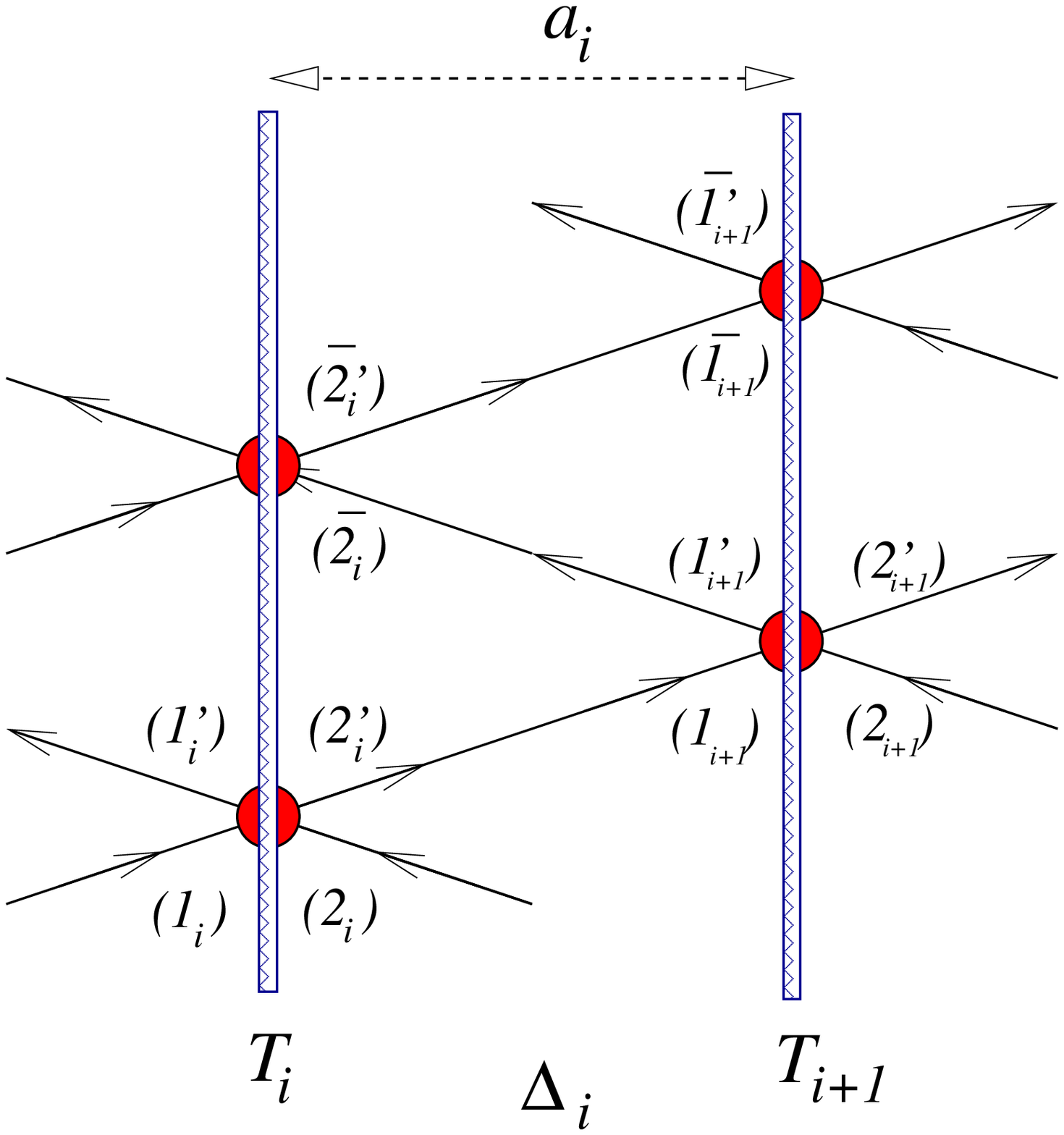}
}
\caption{ 
\hspace{0.1\columnwidth} (a) \hspace{0.45\columnwidth} (b) \newline
The multi-layer structure consisting of $N$ superconducting layers
surrounded by $2$ half-infinite superconductors. The order parameters
are $\Delta_{0}, \cdots, \Delta_{N+1}$, the interface transparencies
are $T_{1}, \cdots , T_{N+1}$, and the thicknesses of the layers
$a_{1}, \cdots , a_{N}$. (a) Scattering on the $i$'th interface with
the transparency $T_{i}$. The angle of the trajectory is $\Theta$.
(b) The zig-zag trajectory inside of $i$'th superconducting layer with
the order parameter $\Delta_{i}$. Due to the translational symmetry
are certain parts of the trajectory equivalent, e.g. $1
\leftrightarrow \bar{1}$, $2 \leftrightarrow \bar{2}$, $\cdots$.  }
\label{fig:multiLayerSetup} \end{figure}

In the quasiclassical technique,, the current density $\mathbf{j}$
reads \cite{LarOvc}:
\begin{equation}
\mathbf{j}  = \int d\varepsilon 
\frac{d \Omega_{\mathbf{n}}}{4 \pi} \;
\mathbf{j}_{\mathbf{n}}(\varepsilon) 
\left( 1 - 2 n(\varepsilon) \right) \; ,
\label{mM01}
\end{equation}
where $\varepsilon$ is energy and $\mathbf{n}$ a unit vector which
shows the direction of the momentum, $n(\varepsilon )$ is the
distribution function, and the angular and energy resolved current
$\mathbf{j}_{\mathbf{n}}(\varepsilon)$ is found as
\begin{equation}
\mathbf{j}_{\mathbf{n}} (\varepsilon) = v_{F} 
\mathbf{n} \; \text{Re} \left( g^{R}_{\mathbf{n}}  ( \varepsilon
)\right)_{11} \; ,
\label{mM02}
\end{equation}
where $v_{F}$ is the Fermi velocity and $g^{R}_{\mathbf{n}}$ is the
retarded quasiclassical Green's function (see e.g. Ref.\cite{SheOza00}
for notation).  For a given energy $\varepsilon$ and a parallel
momentum $p_ {||}$, the $x$ component of the current along the $x$
axis is sum of the contributions $j_{\mathbf{n}_{1}} +
j_{\mathbf{n}_{1'}}$, where $( \mathbf{n}_{1} )_{x} = \cos \Theta $
and $( \mathbf{n}_{1'} )_{x} =- \cos \Theta$ with $\Theta$ being the
angle of the trajectory defined as $\sin \Theta = p_{||} / p_{F}$.  In
order to compare the current with Eq.~(\ref{2kc}), we change the
integration in Eq.~(\ref{mM01}) as follows: $\int d\Omega_{\mathbf{n}}
\leftrightarrow \int dp_{||}/p_{F}$ and $\int_{-\infty}^{\infty}
d\varepsilon \leftrightarrow 2 \int_{0}^{\infty} d\varepsilon$. The
corresponding partial current then reads
\begin{eqnarray}		
j_{p_{||}} ( \varepsilon ) = v_{F} \frac{p_{||}}{p_{F}} \;\text{Re}
\left( g^{R}_{\mathbf{n}_{1}} -g^{R}_{\mathbf{n}_{1'}} \right)_{11} \;
, \label{mM03} \\ \mathbf{j} = \int \limits_{0}^{\infty} d
\varepsilon \int \limits_{0}^{\infty} \frac{dp_{||}}{p_{F}} \;
j_{p_{||}} ( \varepsilon ) \left( 1 - 2 n(\varepsilon) \right) \; ,
\label{mM03a} 
\end{eqnarray} 
where the Green's functions are taken at the same space point, e.g. at
the interface shown on Fig.\ref{fig:multiLayerSetup}a.

To evaluate the Green's function $g$ ($= g^{R}$), we use the method of
Ref. \onlinecite{SheOza00,OzaShe01} and express the $2\times 2$ matrix
$g$ via two-component ``wave functions'' $\phi_{\pm}$ on classical
trajectories:
\begin{equation}
g = \frac{2}{N} \phi_{+} \overline{\phi}_{-} -1 
\quad ; \quad
N = \overline{\phi}_{-} \phi_{+} \; , 
\label{mM04}
\end{equation}
where $\overline{\phi} = -i \phi^{T} \tau_{y}$. These amplitudes obey
Andreev-like equation on classical trajectories (see
\cite{SheOza00,OzaShe01})).  The index $\pm$ denotes solutions with
different asymptotic behaviour: the amplitude $\phi_{+} \rightarrow 0$
for the trajectory coordinate $x$ going to +infinity: $x \rightarrow
\infty$, and $\phi_{-} \rightarrow 0$ for $x \rightarrow -\infty$.

The Andreev equation needs a boundary condition when the trajectory
hits an interface and ballistic pieces of trajectories are tied by a
``knot'', see Fig.\ref{fig:multiLayerSetup}a.  The boundary condition
can be formulated via the transfer matrix $\mathsf{M}_{j' \leftarrow
j}$ \cite{SheOza00},
\begin{equation}
\phi_{j'} = \mathsf{M}_{j' \leftarrow j} \phi_{j} \; .
\label{mM05}
\end{equation}
which relates the wave function on the incoming trajectory $j$ to that
on the outgoing trajectory $j'$; here and below, $j$ denote incoming
channels, whereas $j'$ is the outgoing reflected one, e.g. $j=1_{i}$
and $j' = 1_{i}'$ on Fig.\ref{fig:multiLayerSetup}a.  The ``knot''
transfer matrix $\mathsf{M}_{j' \leftarrow j}$ is expressed in terms
of ``across knot'' Green's function $g_{k' \bullet k}$ which depends
on the amplitudes $\phi_{\pm}$ in the channels on the other side of
the interface \cite{SheOza00}
\begin{eqnarray}
\mathsf{M}_{1_{i}' \leftarrow 1_{i}} &=& \frac{1+R}{2r^{*}} 
\left( 1 - \frac{T}{1+R} g_{2_{i}' \bullet 2_{i}} \; \right) \; ,
\label{mM06} \\
g_{2_{i}' \bullet 2_{i}} &=& 
 \frac{2}{N} \phi_{(2_{i}')+} \overline{\phi}_{(2_{i})-} -1 
\; , 
\label{mM06a}
\\
N &=& \overline{\phi}_{(2_{i})-} \phi_{(2_{i}')+} \; ,
\label{mM06b}
\end{eqnarray}
where $T$ and $R=|r|^{2}$ are the transmission and the reflection
probabilities. The indices $1_{i}$ and $2_{i}$ refer to the channels
on the other side of an interface, see Fig.\ref{fig:multiLayerSetup}a.
According to Eq.(\ref{mM06a}) the ``plus'' amplitude $\phi_{+}$
is needed only in the outgoing channels and the ``minus'' amplitude
$\phi_{-}$ in the incoming ones.

In order to find the Green's function in one of the external channels
(channels which lead to the infinity) we need to calculate the
amplitudes $\phi_{(j)+}$ and $\phi_{(j')-}$ in the channels inside of
the structure.  To find these amplitudes it is convenient to introduce
the total transfer matrix $\mathcal{M}$.  If one
considers the periodic zig-zag trajectory inside of $i$'th layer, see
Fig.\ref{fig:multiLayerSetup}b the $\mathcal{M}$ is defined as the
operator connecting the corresponding parts, for example the
trajectories $(2_{i})$ and $(\bar{2}_{i})$ or $(2_{i}')$ and
$(\bar{2}_{i}')$.  The total transfer matrices in the $i$'th layer
shown on Fig.\ref{fig:multiLayerSetup}b read
\begin{eqnarray}
\mathcal{M}_{\bar{2}_{i} \leftarrow 2_{i}} &=& 
\mathsf{U}_{i} \mathsf{M}_{1_{i+1}' \leftarrow
1_{i+1}} \mathsf{U}_{i} \mathsf{M}_{2_{i}' \leftarrow 2_{i}} \; , 
\label{mM08a} \\
\mathcal{M}_{\bar{2}_{i}' \leftarrow 2_{i}'} &=& 
\mathsf{M}_{\bar{2}_{i}' \leftarrow
\bar{2}_{i}} \mathsf{U}_{i} \mathsf{M}_{1_{i+1}' 
\leftarrow 1_{i+1}} \mathsf{U}_{i} \; ,
\label{mM08b}  \\
\mathsf{U}_{i}  &=& \openone \cos \frac{\xi_{i} a_{i}}{v_{F} \cos \Theta} +
i \hat{g}_{0} \sin \frac{\xi_{i} a_{i}}{v_{F} \cos \Theta} \; ,
\label{mM08c}  \\
\hat{g}_{0} &=& \frac{1}{\xi_{i}}
\left(
\begin{array}{cc}
\varepsilon & -\Delta_{i} \\
\Delta_{i}^{*}  & -\varepsilon
\end{array}
\right) \; ,
\label{mM08d}
\end{eqnarray}
where $\mathsf{U}_{i}$ is the propagator across the $i$'th layer
\cite{SheOza00}, $\xi_{i}^{2} = \varepsilon^{2} - | \Delta_{i}|^{2} ,
\text{Im} \xi > 0$ and $\mathsf{M}_{j' \leftarrow j}$ is the ``across
knot'' transfer matrix. The transfer matrices
$\mathcal{M}_{\bar{1}_{i+1} \leftarrow 1_{i+1}}$ and
$\mathcal{M}_{\bar{1}_{i+1}' \leftarrow 1_{i+1}'}$ are found in the
same way.  As shown in Ref.\cite{SheOza00} the quasiclassical Green's
function is found as
\begin{equation}
g_{j} = \frac{
\mathcal{M}_{\bar{j} \leftarrow j} - 
\openone \frac{1}{2} \text{Tr} \; \mathcal{M}_{\bar{j}
\leftarrow j} 
}{
\sqrt{\left(
\mathcal{M}_{\bar{j} \leftarrow j} - 
\openone \frac{1}{2} \text{Tr} \; \mathcal{M}_{\bar{j}
\leftarrow j} 
\right)^{2}}
} \; .
\label{mM07}
\end{equation}
For example if we want to calculate Green's function in the channel
$(2_{i})$ we first find the total transfer matrix
$\mathcal{M}_{\bar{2}_{i} \leftarrow 2_{i}}$ connecting $\phi$'s in
channels $(2_{i})$ and $(\bar{2}_{i})$ and then we calculate
$g_{(2_{i})}$ from Eq.(\ref{mM07}).

When the quasiclassical Green's functions $g_{(2_{i})}, g_{(2_{i}')},
g_{(1_{i+1})}$ and $g_{(1_{i+1}')}$ are known in each layer, one can
invert Eq.(\ref{mM04}) and calculate the amplitudes $\phi_{(2_{i})-},
\phi_{(2_{i}')+}, \phi_{(1_{i+1})-}$ and $\phi_{(1_{i+1}')+}$:
\begin{eqnarray}
\phi_{(j')+} &=&  \left(
\begin{array}{c}
1 - (g_{j'})_{22} \\
(g_{j'})_{21}
\end{array}
\right) \; ,
\label{mM09a}\\
\phi_{(j)-} &=& 
\left(
\begin{array}{c}
- (g_{j})_{12} \\
1 - (g_{j})_{22}
\end{array}
\right) \; .
\label{mM09b}
\end{eqnarray}
Using Eqs.(\ref{mM04}-\ref{mM09b}) we can write down the
following iterative procedure:

\begin{enumerate}
\item
Set all the knot values of the amplitudes $\phi_{-}$ in the incoming
channels and $\phi_{+}$ in the outgoing channels to the bulk values
\begin{eqnarray}
\phi_{(1_{i}')+} &=&  \phi_{(2_{i}')+} =  \left(
\begin{array}{c}
1 \\
\frac{\Delta^{*}_{i}}{\varepsilon + \xi_{i}}
\end{array}
\right) \; ,
\label{mM10a} \\
\phi_{(1_{i})-} &=& \phi_{(2_{i})-} = 
\left(
\begin{array}{c}
\frac{\Delta_{i}}{\varepsilon + \xi_{i}} \\
1
\end{array}
\right) \; ,
\label{mM10b}
\end{eqnarray}
where $\Delta_{i}$ is the order parameter in the
$i$'th layer, and $\xi^{2}_{i} =
\varepsilon^2 - | \Delta_{i}|^{2}$.

\item \label{step1}
Using the values of the amplitudes $\phi_{\pm}$, the ``across-knot''
Green's functions $g_{2_{i}' \bullet 2_{i}}$ and $g_{1_{i+1}' \bullet
1_{i+1}}$ are constructed from Eq.(\ref{mM06a}).

\item 
The ``across-knot'' Green's functions are substituted into
Eq.(\ref{mM06}) to calculate the knot transfer matrices
$\mathsf{M}_{j' \leftarrow j}$, where here and below index $j$ stays
for $1_{i}$ or $2_{i}$ in $i$'th layer.

\item \label{TotalTransfer}
The total transfer matrices $\mathcal{M}_{\bar{j} \leftarrow j}$ and
$\mathcal{M}_{\bar{j'} \leftarrow j'}$ are calculated using
Eqs.(\ref{mM08a}-\ref{mM08b}) in all layers.

\item 
Using the values of the total transfer matrices from step
\ref{TotalTransfer}.  the quasiclassical Green's functions $g_{j}$ and
$g_{j'}$ are found from Eq.(\ref{mM07}).

\item
At this stage the quasiclassical Green's function $g_{j}$ and $g_{j'}$
are known in all layers and the amplitudes $\phi_{(j)-}$ and
$\phi_{(j')+}$ can be evaluated using
Eq.(\ref{mM09a}-\ref{mM09b}).

\item
Continue from point \ref{step1}. until the convergence is reached.
\end{enumerate} 

After the last iteration the internal amplitudes $\phi_{(j)-}$ and
$\phi_{(j')+}$ are known and one constructs the ``knot'' transfer
matrix $\mathsf{M}_{1' \leftarrow 1}$ on the leftmost external
interface. Since $\phi_{(1')+}$ and $\phi_{(1)-}$ are the bulk
superconductor amplitudes (see Eqs.(\ref{mM10a}-\ref{mM10b})) one
calculates also the Green's functions $g_{(1)}$, $g_{(1')}$ and the
partial current in Eq.(\ref{mM03}).

\section{Results} \label{res}

In this section we present results of the calculations for typical
parameters of the multi-layer structure such as distribution of the
order parameter, thicknesses and number layers (barriers), and the
strength of the barriers. Thicknesses of the layers are usually of
order of $v_{F}/\Delta_{0}$ where $\Delta_{0}$ is the gap in the
leftmost layer.  The value $p_{F}= 10^{3} \Delta_{0}/ v_{F}$ is chosen
for the Fermi momentum.

\begin{figure}
\centerline{
\includegraphics[width=0.8\columnwidth]{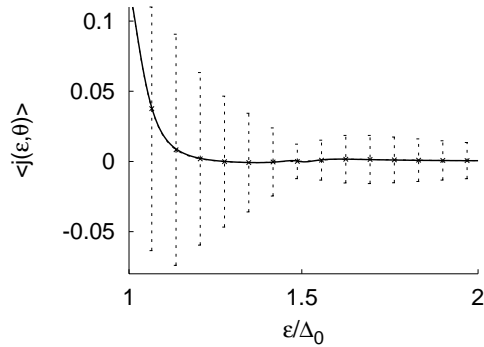}
}
\caption{ 
Angle resolved current in SISIS structure averaged over small range of
$d\theta = 0.09$ as a function of energy. The angle of incidence
$\theta = \pi/4$.  The solid and dashed lines show for the
quasiclassical and coarse-grain BdG current, respectively.  The bars
show the fluctuation of the high-resolution BdG current around its
average value.  The order parameter in the leftmost superconductor is
$\Delta_{0}$. The order parameters in the next layers read $\Delta_{1}
= e^{i\pi/4} \Delta_{0}, \Delta_{2} = i \Delta_{0}$.  The interface
transparencies $T_{1}=0.1, T_{2} = 0.5$.  the thickness of the layer
$a = v_{F}/\Delta_{0}$.  }  
\label{SSSs} \end{figure}

A typical high resolution angular dependence of the current
$j_{BdG}(\theta )$ has been already shown in Fig. \ref{SSS}. As
expected, it does not have any resemblance to the smooth
quasiclassical behaviour. However, after the coarse grain averaging,
{\it i.e.} on a low resolution level, $j_{BdG}$ is in perfect
agreement with quasiclassics, see Fig.\ref{SSSs}.  The ``error-bars''
in Fig.\ref{SSSs} show the mean square fluctuation of the high
resolution current around its coarse-grain average.  In the
double-barrier case, the agreement exists for any angle $\theta $ and
energy $\epsilon $, and for any set of parameters of the structure,
$\Delta $'s and the barrier's strength.

Contrary, noticeable deviations from the quasiclassical solutions are
seen when there are more than two barriers.  Here we present results
only for three-barrier structures, more complicated systems show
qualitatively same features.

\begin{figure}
\centerline{\includegraphics[width=0.8\columnwidth]{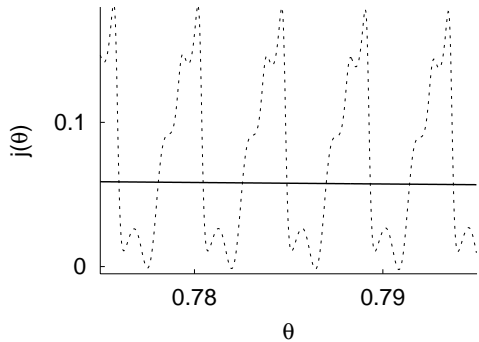}}
\caption{
Angle resolved current in a three-barrier structure SISISIS as a
function of angle $\theta$ as found from the BdG equation.  The order
parameter in the leftmost superconductor is $\Delta_{0}$. The order
parameters in the next layers read $\Delta_{1} = i \Delta_{0},
\Delta_{2} = - \Delta_{0}, \Delta_{3} = -i \Delta_{0}$.  The
thicknesses of the two internal layers $a_{1} = v_{F}/\Delta_{0},
a_{2} = v_{F}/\Delta_{0}$.  The energy $\varepsilon =1.1 \Delta_{0}$.
The interface transparencies $T_{1}=0.1, T_{2} = 0.5, T_{3} = 0.5$.
The solid line corresponds to the quasiclassical current, a constant
in this narrow interval. The dashed line shows the BdG current.
\label{ang155E1.1} } \end{figure}

\begin{figure}
\centerline{
\includegraphics[width=0.48\columnwidth]{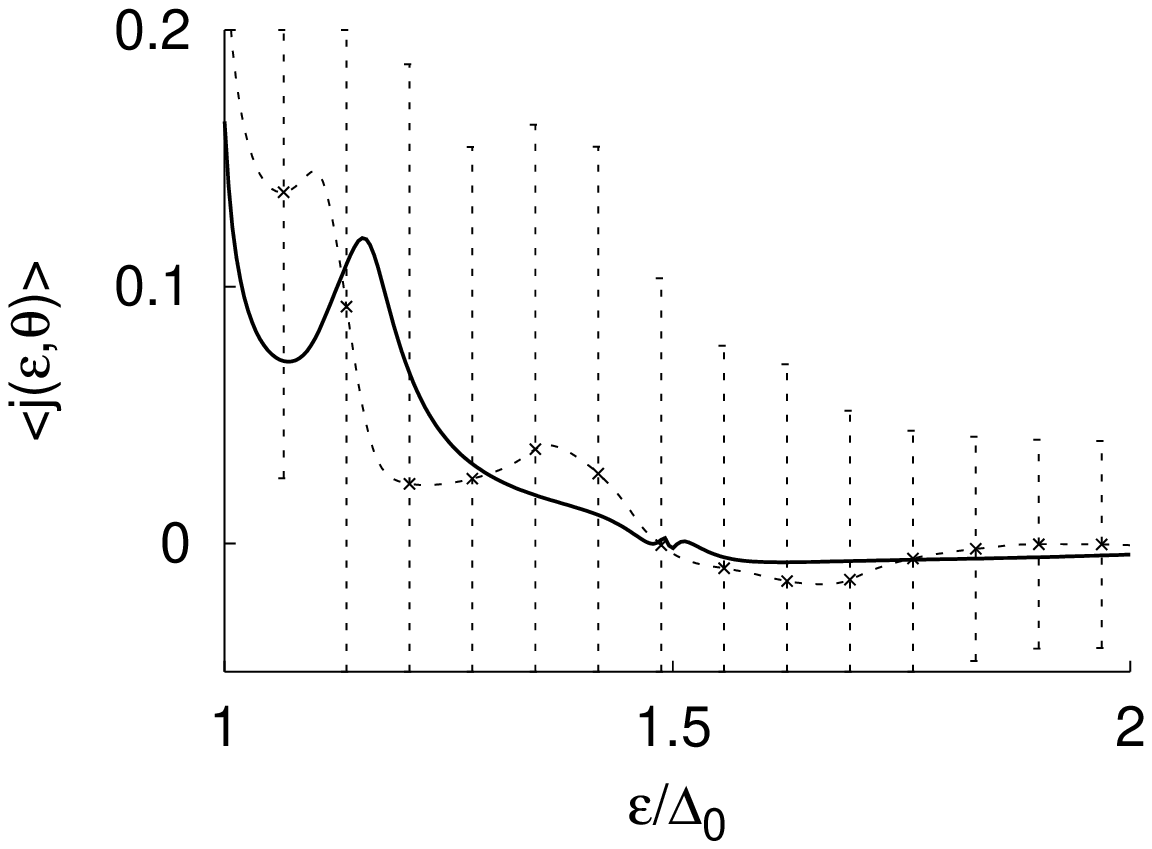}
\hspace{0.05\columnwidth}
\includegraphics[width=0.48\columnwidth]{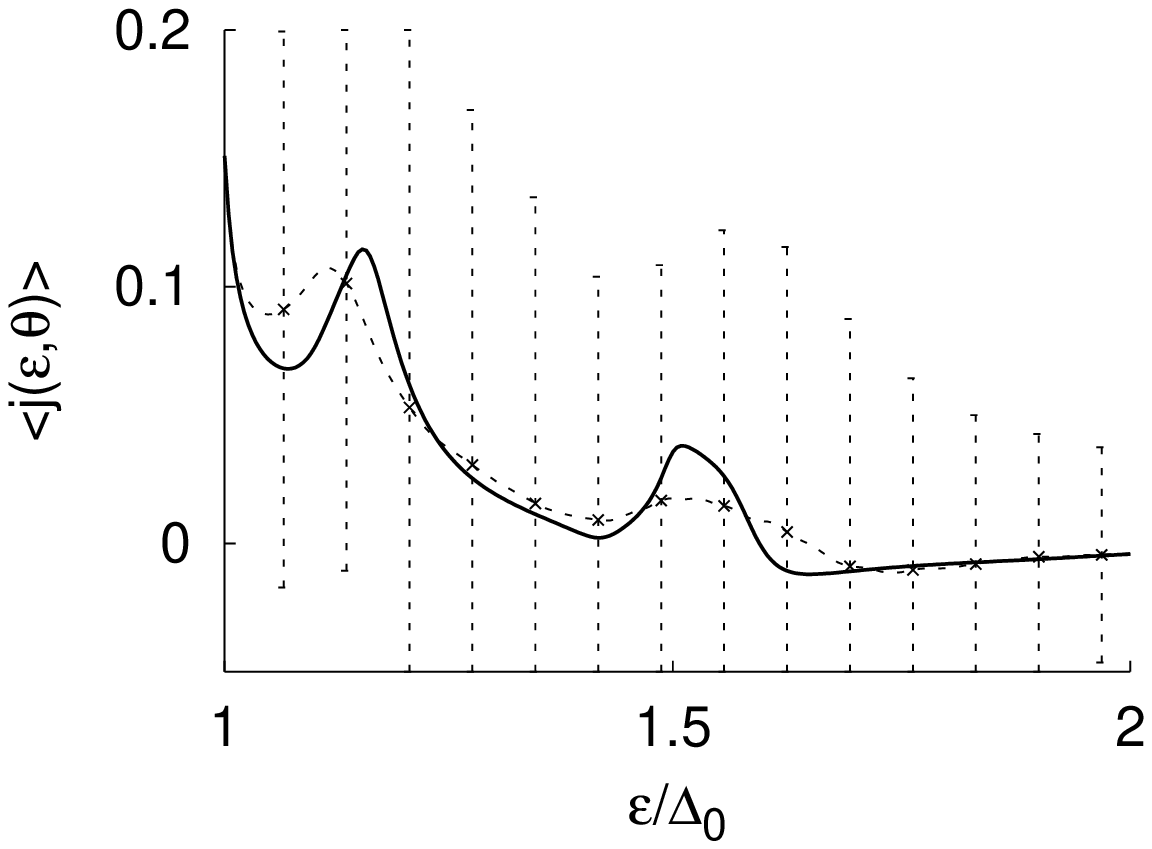}
}
\caption{ 
\hspace{0.1\columnwidth} (a) \hspace{0.45\columnwidth} (b) \newline
Angle resolved current averaged over small range of $d\theta = 0.09$
as a function of energy. The angle of incidence $\theta = \pi/4$.  The
solid line stands for the quasiclassics and the dashed line for the
BdG current. The bars show how the BdG current changes around its
average value.  The order parameter in the leftmost superconductor is
$\Delta_{0}$. The order parameters in the next layers read $\Delta_{1}
= i \Delta_{0}, \Delta_{2} = - \Delta_{0}, \Delta_{3} = -i
\Delta_{0}$.  The interface transparencies $T_{1}=0.1, T_{2} = 0.5,
T_{3} = 0.1$.  (a) Symmetric case: the thicknesses of the two internal
layers $a_{1} = v_{F}/\Delta_{0}, a_{2} = v_{F}/\Delta_{0}$.  (b)
Non-symmetric case: $a_{1} = 0.9 \; v_{F}/\Delta_{0}, a_{2} = 1.1\;
v_{F}/\Delta_{0}$.  }  
\label{arc151sym-arc151asym} \end{figure}

For a three-interface structure, a typical high resolution angle
dependence of $j_{BdG}$ is shown in Fig. \ref{ang155E1.1}.  In
Fig.\ref{arc151sym-arc151asym} we plot coarse-grain BdG-current
together with the quasiclassical curve for slightly different
geometries: Fig.\ref{arc151sym-arc151asym}(a) refers to a symmetric
case when the two internal layers have exactly the same thickness
whereas in \ref{arc151sym-arc151asym}(b) the thicknesses are 10
percent different from each other.  In the both cases, one sees a
clear deviation of the quasiclassical curve from the ``exact'' one,
the deviation lesser in asymmetric geometry.

\begin{figure}
\centerline{
\includegraphics[width=0.48\columnwidth]{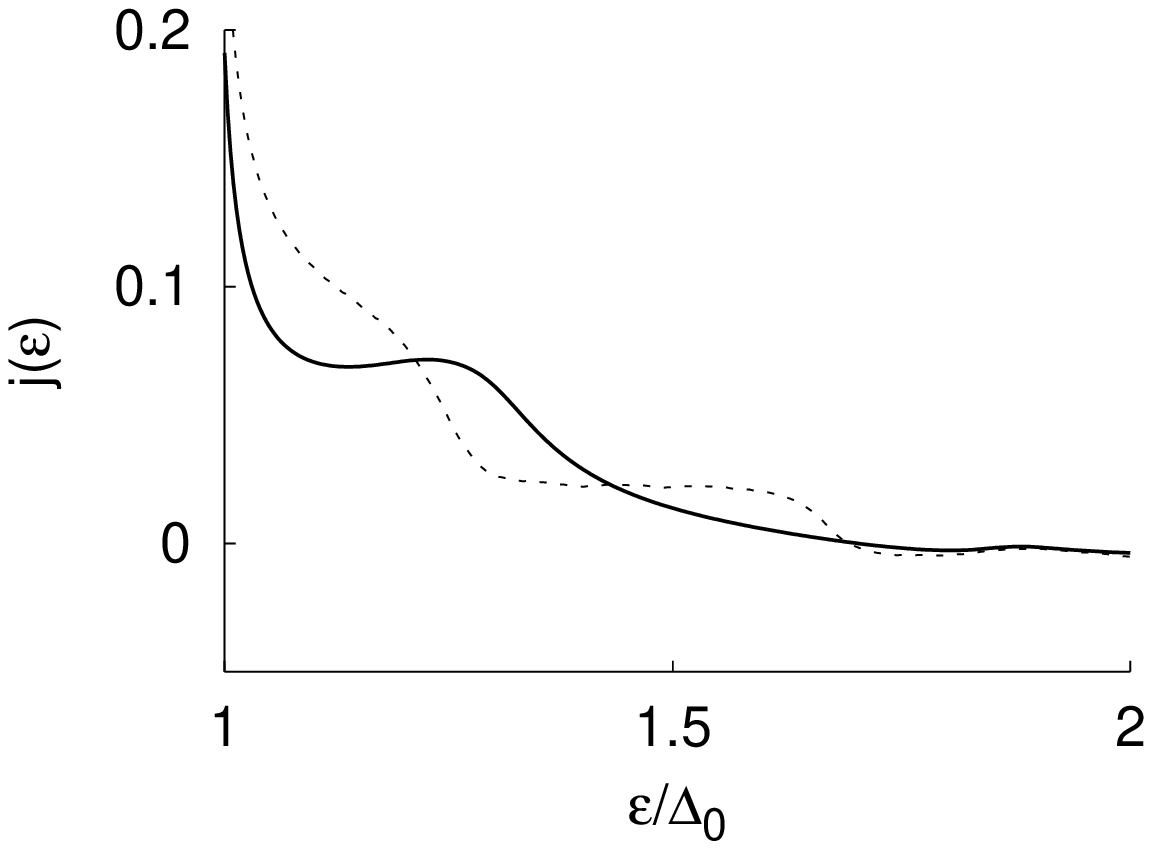}
\hspace{0.05\columnwidth}
\includegraphics[width=0.48\columnwidth]{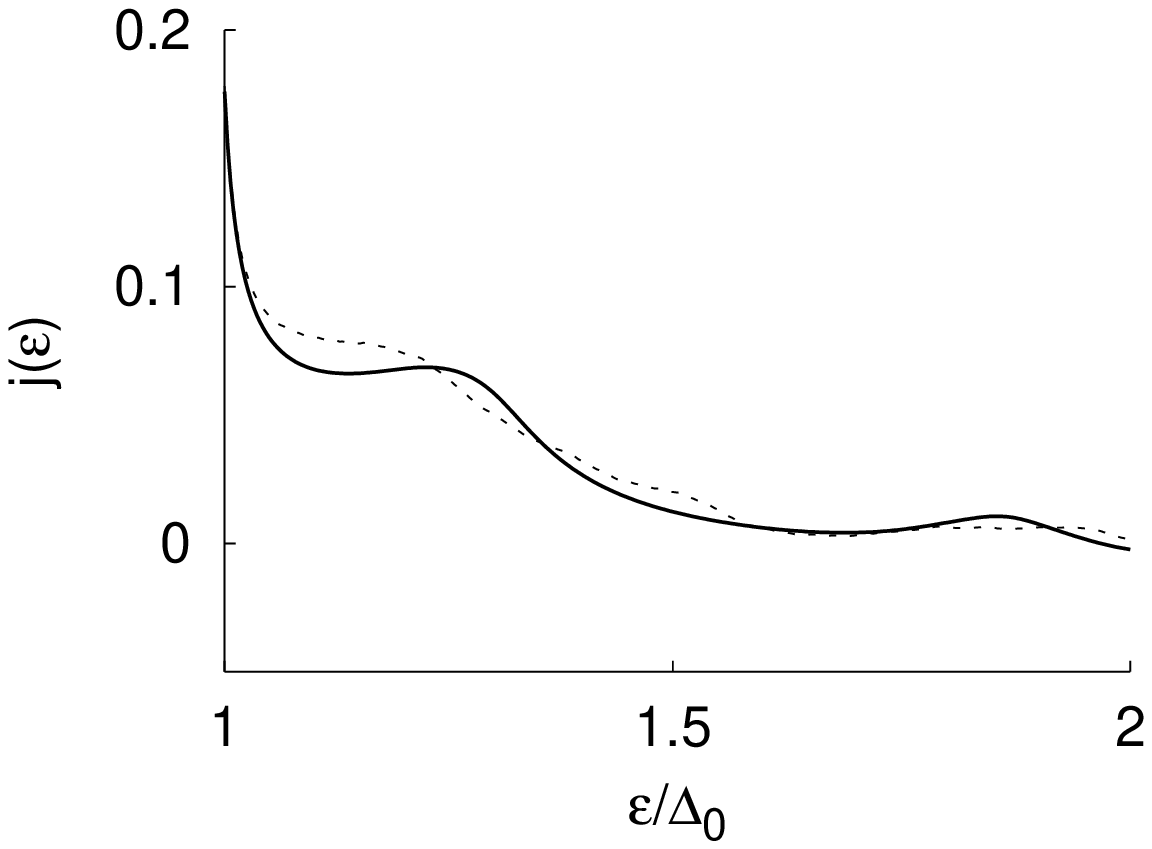}
}
\caption{ 
\hspace{0.1\columnwidth} (a) \hspace{0.45\columnwidth} (b) 
\newline
The total current as a function of energy.  The solid line stands for
the quasiclassical and the dashed line for the BdG current.  The
order parameter in the leftmost superconductor is $\Delta_{0}$. The
order parameters in the next layers read $\Delta_{1} = i \Delta_{0},
\Delta_{2} = - \Delta_{0}, \Delta_{3} = -i \Delta_{0}$.  The interface
transparencies $T_{1}=0.1, T_{2} = 0.5, T_{3} = 0.1$.  (a) Symmetric
case: the thicknesses of the two internal layers $a_{1} =
v_{F}/\Delta_{0}, a_{2} = v_{F}/\Delta_{0}$.  (b) Asymmetric case:
$a_{1} = 0.9 \; v_{F}/\Delta_{0}, a_{2} = 1.1\; v_{F}/\Delta_{0}$.  }
\label{tot151sym-tot151asym} 
\end{figure}

For the same geometries and the order parameters, the energy
dependence of the current integrated with respect to the incident
angle are shown in Fig.\ref{tot151sym-tot151asym}. Disagreement
between the exact and quasiclassical results are clearly seen, again
stronger in the symmetric case.

\section{Discussion}\label{diss}

The results of the previous section clearly show that in some
geometries the quasiclassical theory does not reproduce the ``exact''
results derived from the BdG equation.  The two approaches agree only
qualitatively. The quantitative discrepancy much exceeds the
corrections to the quasiclassical theory of order of $1/ p_{F}a
\sim\Delta / p_{F}v_{F}$ which one might expect. Below, we present our
understanding of physics behind the discrepancy.

\begin{figure}
\centerline{
\includegraphics[width=0.43\columnwidth]{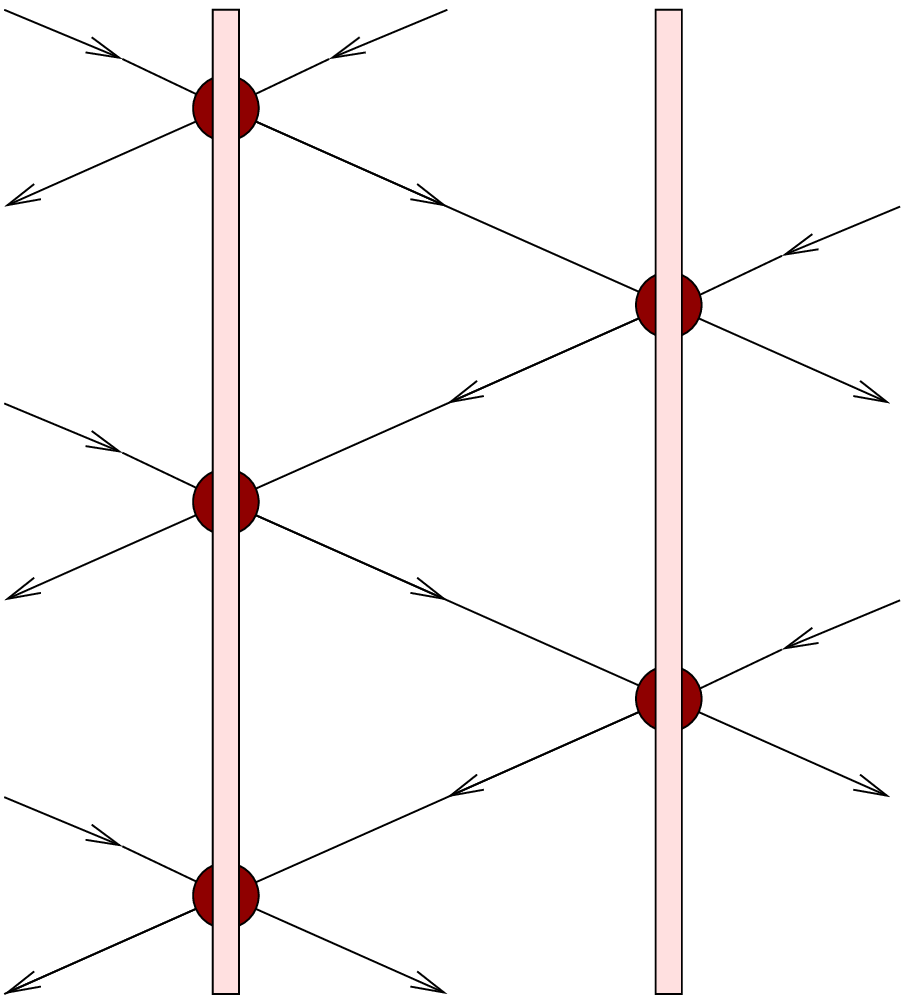}
\hspace{0.05\columnwidth}
\includegraphics[width=0.43\columnwidth]{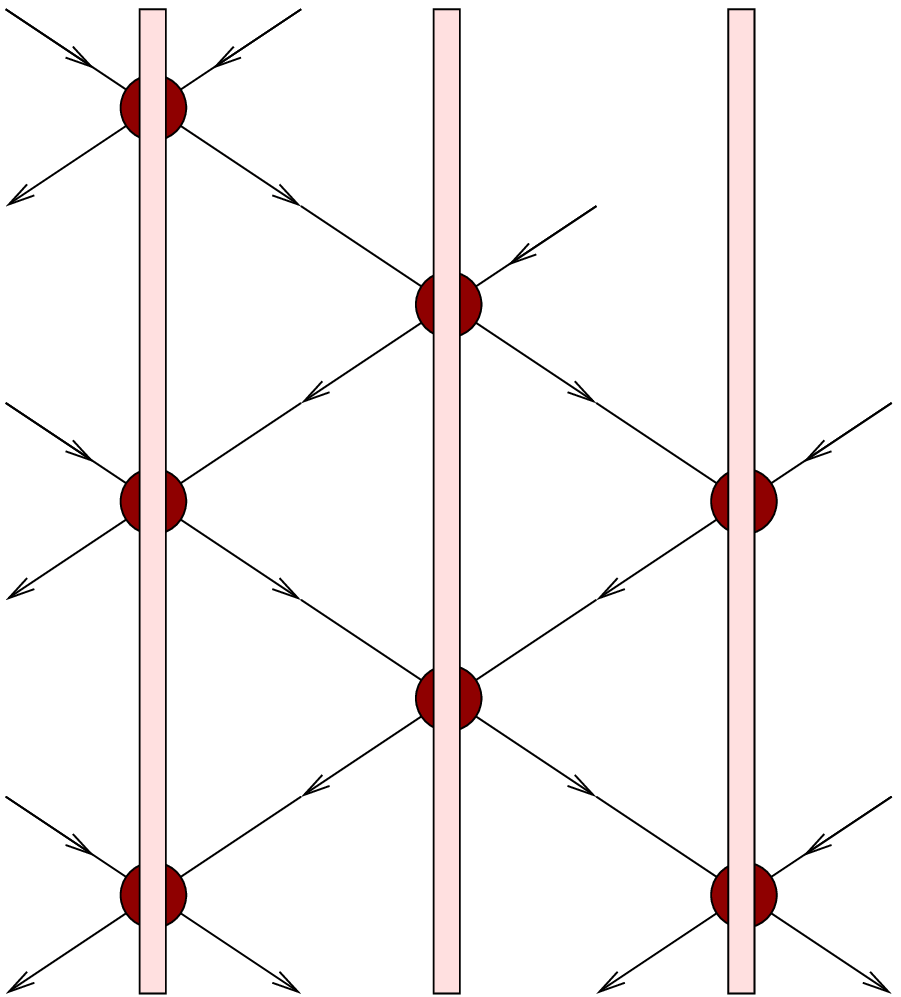}
}
\caption{ 
\hspace{0.1\columnwidth} (a) \hspace{0.45\columnwidth} (b) \newline 
Classical trajectories. The
trajectory is built of ballistic pieces ``tied'' by scattering on the
two interfaces (knots). The arrows show the direction of the momentum.
There are no interfering paths in a double-layer case (a). 
Loops exist in a three-barrier system (b).
}
\label{two-three}
\end{figure}

As in our earlier papers \cite{SheOza00,OzaShe01,OzaShe02}, we ascribe
the failure of the quasiclassical theory to the presence of
interfering paths or, in other words, loop-like trajectories. From
this point, the validity of the quasiclassical theory in the
two-barrier case (see Fig.\ref{SSSs}) is in accordance with our
expectations. Indeed, in this simple geometry, the classical path
shown in Fig. \ref{two-three}a, is effectively one dimensional
(tree-like trajectory in the terminology of Ref. \cite{SheOza00}) in
the sense that there is only one path connecting any two points.  As
discussed in Ref.\cite{SheOza00}, one is then able to factorize the
full propagator $G(x,x')$, $x$ $x'$ labelling the points on the
tree-like trajectory, as $G(x,x')= g(x,x') \exp[i p_{F}{\cal L}_{xx'}]
$, where ${\cal L}_{xx'}$ is the length of the path along the
tree-like trajectory connecting $x$ and $x'$, and $g(x,x')$ is a
slowly varying quasiclassical (2-point) Green's function. In this
case, derivation of the quasiclassical equation does not meet any
difficulty, and the theory gives valid results.

Turning now to the symmetric three-barrier structure, the trajectory
is {\em not} simple tree-like (see Fig.\ref{two-three}(b)): there are
loops and, therefore, interfering paths. Then, one may expect
corrections to the quasiclassical theory, which are not controlled by
the quasiclassical parameter \cite{OzaShe02}.  The origin of the
corrections and its relation to the existence of loops, can be
understood from the following semi-quantitative arguments.

\begin{figure}
\centerline{
\includegraphics[width=0.43\columnwidth]{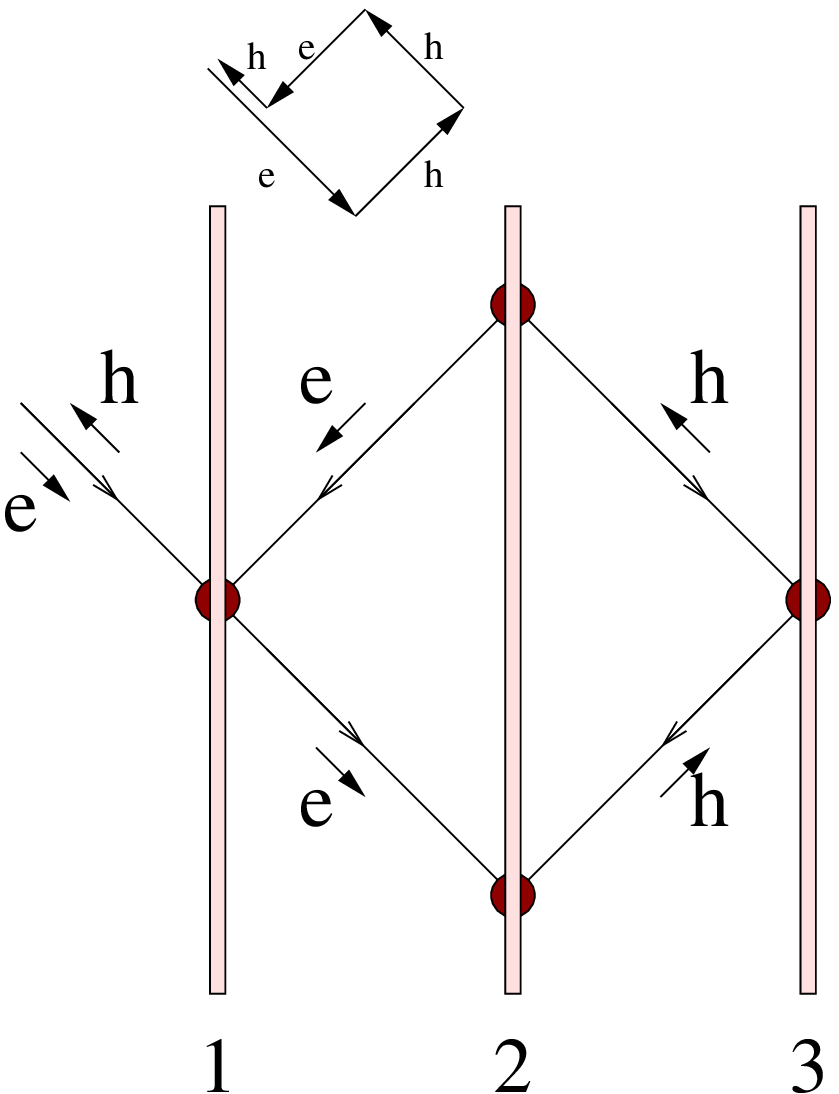}
\hspace{0.05\columnwidth}
\includegraphics[width=0.43\columnwidth]{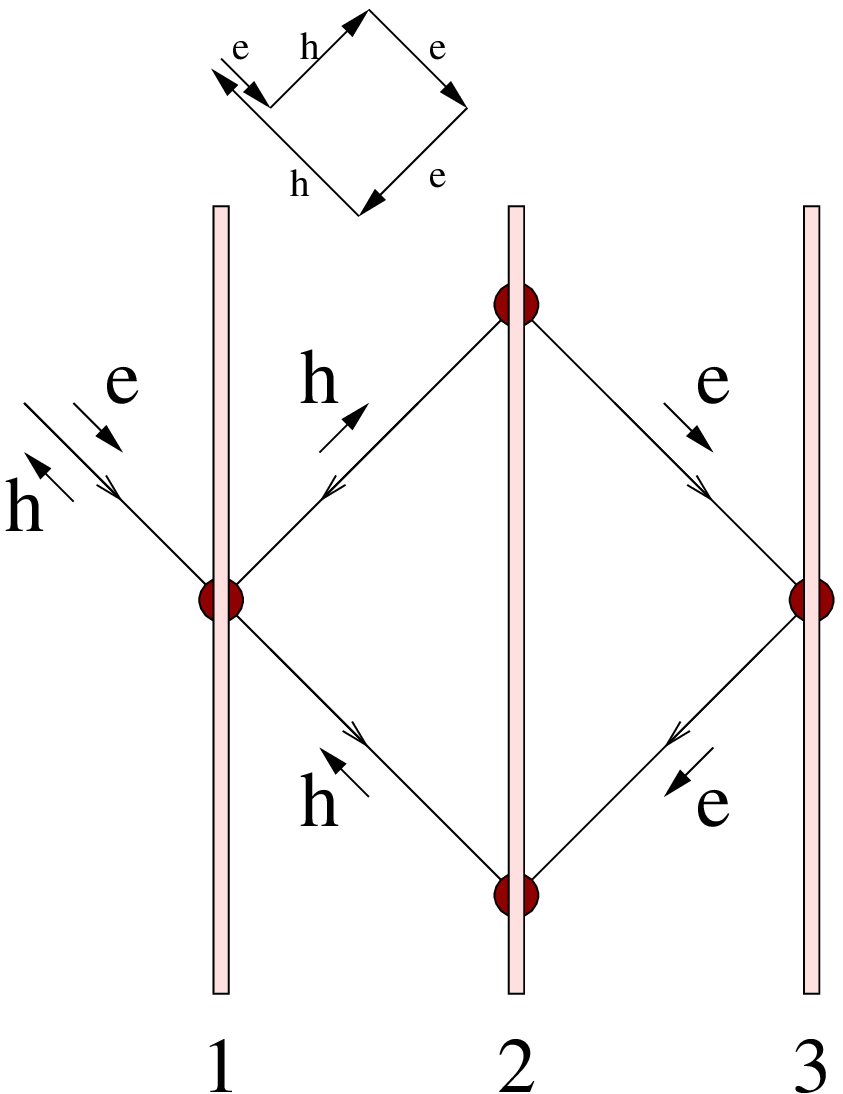}
}
\caption{ 
\hspace{0.1\columnwidth} (a) \hspace{0.45\columnwidth} (b) \newline 
Simplest loop like paths contributing to the Andreev reflection amplitude
in a symmetric double-barrier structure.
The incoming electron comes from the left and is reflected as a hole
after going around the loop  anti-clockwise (a) and clockwise
(b). Arrows on the continuous lines show the direction of momentum.
Letters ``e'' and ``h'' show the character of the excitation --
electron or hole. The arrows under the letters show the direction of
propagation.
}
\label{loops}
\end{figure}

In the BdG-picture the current Eq.~(\ref{2kc}) is expressed via the
element of the S-matrix $S_{12}$ that is the Andreev reflection
amplitude $A_{e} \equiv S_{12}$.  Scattering amplitudes can be
presented as a sum of partial amplitudes, each of which corresponds to
a particular path of the particle.  Among others, there are closed
paths shown in Fig.\ref{loops}.  The contribution
$A_{e}^{\text{loops}}$ of the paths in Fig.\ref{loops}(a) and (b) to
the full Andreev reflection amplitude $A_{e}$ is evaluated in appendix
\ref{ss-matrix},
\begin{equation}
A_{e}^{\text{loops}}  =  {\cal A}\;
\Re \left( 
r_{1} r_{3}^{*} e^{2i \left(\cos \theta p_{F}(a_{1} -a_{2})\right)}
\right) 
\label{nmc}
\end{equation}
where ${\cal A}$ is a coefficient defined in Eq.~(\ref{vmc3}), $a_{1}$
($a_{2}$) is the distance from the barrier 1 to barrier 2 (from 2 to
3); $r_{1}$ and $r_{3}$ are the amplitudes of reflection from the
barrier 1 and 3 respectively. Note that this simplest loop survives
the coarse grain averaging only if $a_{1}\approx a_{2}$. We understand
the larger deviation from quasiclassics seen in the symmetric case
compared with an asymmetric one, as due to the contribution of the
simple loop.

\begin{figure}
\centerline{
\includegraphics[width=0.8\columnwidth]{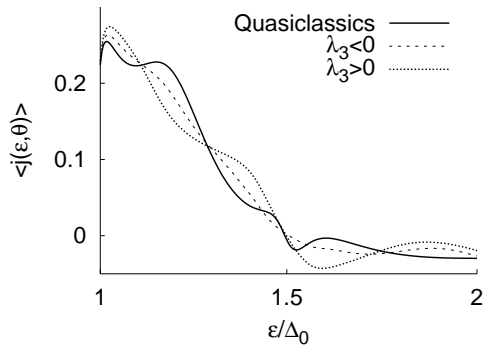}
}
\caption{  
Angle resolved current averaged over small range of $d\theta = 0.09$
as a function of energy. The angle of incidence $\theta = \pi/4$.  The
solid line stands for the quasiclassical current. The other two lines
correspond to $\lambda_{3}$ positive or negative.  The order parameter
in the leftmost superconductor is $\Delta_{0}$. The order parameters
in the next layers read $\Delta_{1} = i \Delta_{0}, \Delta_{2} = -
\Delta_{0}, \Delta_{3} = -i \Delta_{0}$.  The interface transparencies
$T_{1}=0.5, T_{2} = 0.5, T_{3} = 0.5$.  The thicknesses of the two
internal layers $a_{1} = v_{F}/\Delta_{0}, a_{2} = v_{F}/\Delta_{0}$.
} 
\label{arc555Compare}
\end{figure}

\begin{figure}
\centerline{
\includegraphics[width=0.8\columnwidth]{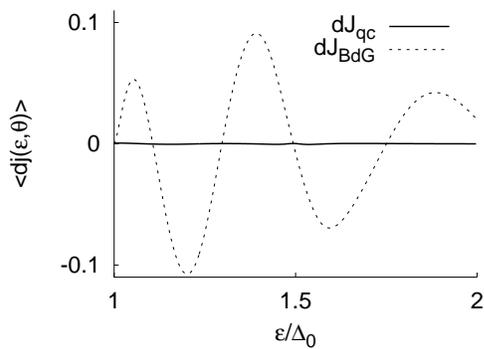}
}
\caption{ 
The differences between currents in SSSS setups differing only by the
thickness of the last layer.  The thicknesses are $a_{1}= a_{2}=
v_{F}/\Delta_{0}$ in the first case and $a_{1}=v_{F}/\Delta_{0} $ and
$a_{2}=1.002 v_{F}/\Delta_{0} $ in the second case.  The order
parameter in the leftmost superconductor is $\Delta_{0}$. The order
parameters in the next layers read $\Delta_{1} = i \Delta_{0},
\Delta_{2} = -\Delta_{0}, \Delta_{3} = -i \Delta_{0}$.  The interface
transparencies $T_{1}=0.5, T_{2} = 0.5, T_{3} = 0.5$.  
}
\label{diff555} 
\end{figure}

The existence and importance of this contribution can be checked
exploiting the fact that it is sensitive to the phase of the
reflection coefficients and the length of the path on the scale of $1/
p_{F}$.  In Fig.\ref{arc555Compare}, we show the current for different
signs of the barrier strength $\lambda_{3}= \pm |\lambda_{3}|$,
changing the phase of $r_{3}= - i \lambda_{3}/ (p_{Fx} + i
\lambda_{3}))$ but leaving the reflection probability intact.  In
Fig.\ref{diff555}, we plot the change of the exact and quasiclassical
currents upon tiny variation of the right layer thickness
(corresponding to $\pi $-change of the Fermi phase factor in
Eq.~(\ref{nmc})).  While quasiclassical current remains intact,
clearly seen changes are observed in the exact current with the order
of magnitude consistent with Eq.~(\ref{nmc}).

To avoid confusion, we remind that we deal with coarse-grain averaged
currents, and therefore the observed sensitivity to the thickness and
the phase of reflection has nothing to do with the size effects (due
to the commensurability of the thickness and the Fermi wave length)
well known in the normal case. We note also, that the loops in
Fig.\ref{loops} do not exist in the normal state because the
electron-hole conversion on the interface 2 would not be possible.

We assert that the loop contribution Eq.(\ref{nmc}) is chiefly
responsible for the deviations from the quasiclassical theory.
Obviously, this contribution cannot be grasped by quasiclassics since
$A_{e}^{\text{loops}}$ is sensitive to the phase of the reflection
amplitudes $r_{1}$ and $r_{3}$, whereas the quasiclassical boundary
condition \cite{Zai84,Esc00,SheOza00} contain only the probabilities
$|r|^{2}$ and $|t|^2$.  This is our argument supporting our
interpretation of the numerical results.  Note that the interpretation
is consistence with the observation that the deviation from
quasiclassics are significantly smaller in the asymmetric case
Fig.\ref{tot151sym-tot151asym}: There, the simple loops are absent and
the deviations from the conventional quasiclassics come from higher
order loops (like that analysed in \cite{OzaShe02}).


\section{Conclusions}\label{conc}

In this paper we have examined the applicability of quasiclassical
theory for description of multiple interface scattering by comparing
quasiclassical solutions with ``exact'' ones, extracted from
coarse-grain averaged solutions to the Bogoliubov - de Gennes
equation.  We see that the two approaches agree in simple geometries
(one or two interfaces) but show noticeable discrepancy when three or
more interfaces are present.  This gives an example of a physical
system where quasiclassical technique fails to give quantitative
description with its expected accuracy $\sim 1/ p_{F}\xi_{0}$.  As we
understand it, the failure of the quasiclassical theory occur when
classical trajectories form closed loops (interfering paths) after
sequential reflection and transmission accompanied by electron-hole
conversion. This gives additional support to the point of view of
Ref.\cite{SheOza00} that the derivation of the quasiclassical
technique is possible only under the assumption of a simply connected
topology, tree-like, of classical trajectories.

The main goal of this paper has been to demonstrate the existence of
noticeable deviations from the quasiclassical theory in conditions
where one might expect it to give fully reliable results.  For this
purpose we have chosen the simplest ``exact'' method, the Bogoliubov -
de Gennes approach, where the superconductivity enters via the
mean-field order parameter $\Delta (\bm{r})$ and scattering due to
either impurity or surface roughness is not included.  It is now time
to discuss to what extent our results are sensitive to the
simplifications.

The truly ``exact'' theory of (phonon-mediated) superconductivity, for
which the quasiclassical technique is an approximation, is the set of
Gor'kov-Eliashberg equations for disorder averaged Green's function
${\cal G}_{\varepsilon }$\cite{AbrGorDzy63}. The equation of motion,
$\left(\varepsilon - {\cal H}_{\varepsilon }\right){\cal
G}_{\varepsilon }=1$, contains the operator ${\cal H}_{\varepsilon }$
which has the same structure as the Bogoliubov - de Gennes Hamiltonian
Eq.~(\ref{1}), with the order parameter replaced by the anomalous part
of the electron-phonon self-energy.  Apart from the $\varepsilon
$-dependence of ${\cal H}_{\varepsilon }$, the only qualitative
difference is that the ${\cal H}_{\varepsilon }$ has a non-Hermitian
part coming from the self-energies. The latter accounts for the
impurity and electron-phonon scattering and, correspondingly, for a
finite life time of the excitation with a given momentum,
$\tau(\varepsilon )$ -- the scattering-out time.  Due to the
similarity of the above operators, the interference contribution to
observables obtained with the help of the Green's function technique
or simple minded mean-field Bogoliubov - de Gennes equation, would
give same results.  A finite life time is the only important feature
missing in the Bogoliubov - de Gennes approach, and below we discuss
its role.

Clearly, the interference of waves having travelled different paths
occurs if only the decay length is not too small compared with the
path lengths.  By virtue of the optical theorem, the waves
corresponding to the ballistic trajectories decays on the distance
$\sim v_{F} \tau$.  In practice, $\tau$ is controlled by the bulk
impurity scattering, so that the loops in Fig. \ref{loops} may
contribute only if the interlayer distances are less or of order of
the impurity mean path.  As far as interface imperfection is
concerned, short-range surface roughness is expected to play the role
similar to that of the bulk impurity scattering: Although the
interface reflection becomes partially diffusive, the coherent
specular component, with the intensity proportional to the Fuchs's
parameter ${\cal P}$, is finite \cite{Fuc38,GanLev87}. Thus, the
picture of trajectories as that shown in Fig. \ref{loops} remains
meaningful as well as Eq.~(\ref{nmc}) if corresponding attenuation
factors are inserted.  We see that although microscopic roughness and
bulk scattering suppress the interference effect, it survives disorder
averaging if the disorder is not too strong.

When a long-range roughness is present, that is the layer thicknesses
are slowly varying, the global value of the loop contribution in
Eq.~(\ref{nmc}) averages to zero (if the thickness modulation exceeds
$\lambdabar_{F}$).  However, the interference of paths forming the
loops is sensitive mainly to the local value of the thicknesses, so
that the interference effects can be seen in the spatial fluctuations
of the local current, which is a measurable quantity.  In the
quasiclassical theory, the roughness would reveal itself only of the
thickness modulation is somehow comparable with the coherence length.
Since the loops are sensitive to variations of geometry on the scale
of $\lambdabar_{F}$, their contribution to the fluctuations is
expected to be much larger than that in the quasiclassical theory.

We see that the deviation from the quasiclassical theory due to the
interference effects, although most pronounced in the idealized model
exploited in the paper, are observable in realistic conditions if the
disorder is not too strong.  After disorder averaging, the
interference contribution is observable if the mean free path exceeds
the interlayer distance and the rough interfaces have not too small
coefficient of the specular reflection.  Detail analysis of the loop
contribution to mesoscopic fluctuations is beyond the scope of the
paper.

Apart from disorder, the interference contribution may be suppressed
by energy integration. Indeed, the integration with the Fermi-Dirac
distribution function corresponding to the temperature $T$ is
equivalent to the Matsubara summation, that is energy variable
$\varepsilon$ assumes discrete imaginary values, multiples of $i \pi
T$.  Then, the waves decay on the length $\sim \xi_{0}$, and,
consequently, the loops contribute to equilibrium properties only if
the paths are shorter than the coherence length.

Summarizing, we have shown that the quasiclassical technique fails in
geometries where classical trajectories form closed loops. In
particular, this happens when it is applied to the Josephson
multi-layer structure with number of semi-transparent interfaces
larger than 2.  This conclusion does not undermine the conventional
quasiclassical technique but only limits its applicability in some
special geometries where the interference of classical paths cannot be
neglected.  Within the quasiclassical approach, the interference can
be incorporated into the theory with the help of the method suggested
in Ref. \onlinecite{OzaShe02}.

\begin{acknowledgments}
We are grateful to D. Rainer for discussions and critical remarks.
This work was supported by the University of Ume{\aa}.
\end{acknowledgments}

\appendix

\section{The transfer and scattering matrices}\label{trans}

The
Bogoliubov - de Gennes equation is a second-order differential
equation. In the one-dimensional case considered here, it can be
reduced to the first order equation for an
``extended'' wave functions $\Psi $ which is built of the wave
function $\psi $ and its derivative $\hat{p}\psi$,
\begin{equation}
\Psi(x) = \left(
\begin{array}{c}
\psi \\
\hat{p}\psi
\end{array}
\right)_{x}
 \quad,\quad  \hat{p}= -i \frac{d }{dx}
\; .
\label{Mjnb}
\end{equation}
Since $\psi $ has two components, the extended wave function is a
4-component column.

In terms of $\Psi $,
the quasiparticle current density
Eq.~(\ref{7ic})
reads
\begin{equation}
j^{qp}=  
\frac{1}{2m}\Psi ^{\dagger} \hat{\tau_{z}}  \hat{\sigma}_{x} \Psi 
\label{sjc}
\end{equation}
where $\sigma_{x}$ is the Pauli matrix operating in the space $\psi -
\hat{p}\psi $, and, as before, $\hat{\tau_{z}} $ acts in the $u-v$ space. 

The charge current Eq.~(\ref{ajc}) can be found as 
\begin{equation}
\frac{1}{e} J = \frac{1}{2m}\Psi ^{\dagger}   \hat{\sigma}_{x} \Psi
\; . 
\label{tjc}
\end{equation}

The extended wave function corresponding to the plane waves
 in Eq.~(\ref{5}) have the form
\begin{equation}
\Psi_{\nu \sigma}(x) = \sqrt{\frac{m}{\pi }}\; e^{i \sigma p_{\nu} x}\;
\Phi_{\nu \sigma}
\label{wjc}
\end{equation}
where the 4-component amplitudes $\Phi_{\nu \sigma}$ may be taken as
\begin{equation}
\Phi_{\nu \sigma} = {1 \over \sqrt{2 p_{\nu}}} 
\left(
\begin{array}{r}
\psi_{\nu} \\ \sigma  p_{\nu}\; \psi_{\nu}
\end{array}
\right) \; ,
\label{ujc}
\end{equation}
or in a more concise form,
\begin{equation}
\Phi_{\nu \sigma} =\psi_{\nu} \otimes
\phi_{\sigma, p_{\nu}}
\label{Mlnb}
\end{equation}
where
\begin{equation}
\phi_{\sigma ,p_{\nu}} = {1 \over \sqrt{2 p_{\nu}}} 
\left(
\begin{array}{c}
1 \\ \sigma p_{\nu}
\end{array}
\right) \; .
\label{vjc}
\end{equation}
The amplitudes $\Phi_{\nu \sigma}$ are normalized to the probability
flux Eq.~(\ref{sjc}) equal to $ 1/2m$ (for $E> |\Delta|$);
accordingly,  the plane waves in Eq.~(\ref{wjc}) are normalised to
the $\delta$-function of energy.

The conjugated amplitudes $\Phi^{\ddagger}$ defined as
\begin{equation}
{\Phi}_{\nu \sigma}^{\ddagger} = 
{\psi}_{\nu}^{\ddagger} \otimes {\phi}_{\sigma, p_{\nu}}^{\ddagger}
\quad \text{where}\quad 
{\phi}_{\sigma, p_{\nu}}^{\ddagger} =
\sqrt{p_{\nu} \over 2}
\left(
1 , {\sigma \over p_{\nu}}
\right) \; ,
\label{Mmnb}
\end{equation}
($\phi_{\sigma ,p}^{\ddagger} = 
\sigma \phi_{\sigma ,p}^{T} \hat{\sigma}_{x}$)
satisfy
the following
orthogonality and completeness relations
\begin{equation}
{\Phi}_{\nu' \sigma'}^{\ddagger}\Phi_{\nu \sigma} = \delta_{\nu' \nu}
\delta_{\sigma' \sigma}  
\quad , \quad
\sum \limits_{\nu, \sigma} \Phi_{\nu \sigma} 
{\Phi}_{\nu, \sigma}^{\ddagger} = \hat{\openone} \; .  
\label{Mnnb}
\end{equation}

Due to the orthogonality property, $\Phi_{\nu ,\sigma }^{\ddagger}$
projects a general superposition $\Psi (x)$ to the plane wave
$\Psi_{\nu ,\sigma}(x)$. Using this argument, one constructs
the evolution operator $\hat{U}_{a}$, a $4\times 4$ matrix, which
relates the wave functions at the points $x$ and $x+a$, $\Psi (x)=
\hat{U}_{a} \Psi (x+a)$,
\begin{equation}
\hat{U}_{a}= 
\sum \limits_{\nu, \sigma} e^{-i \sigma p_{\nu} a}
\Phi_{\nu \sigma} {\Phi}_{\nu \sigma}^{\ddagger}  \; .
\label{xjc}
\end{equation}

In the model where the potential barrier $V(x)$ separating the layers
is a $\delta -$function, $V(x)= \frac{\lambda }{m} \delta (x)$, the
two-component wave function $\psi (x)$ is continuous at $x=0$, whereas the
derivatives suffer 
the jump: $\psi' |_{0-}^{0+}= \lambda \psi (0)$.
In terms of the extended wave function $\Psi $ Eq.~(\ref{Mjnb}),
the  interface matching condition reads 
\begin{equation}
\Psi_{0^{-}} = {\cal D} \Psi_{0^{+}} \quad ; \quad 
{\cal D} = 
\left(
\begin{array}{cc}
1 & 0 \\
2i \lambda & 1
\end{array}
\right)  \; .
\label{Mknb}
\end{equation}
It is implied in Eq.~(\ref{Mknb})
that each matrix element of ${\cal D}$ is multiplied by the unit
matrix in the $u-v$ space so that ${\cal D}$ is actually a $4 \times4$
\cite{generalD}. 

The transfer matrix, 
${\cal M}$, 
relates the
extended wave function, $\Psi^{(L)}$, on the left side of the
multi-layer structure to that on the right side $\Psi^{(R)}$:
\begin{equation}
\Psi^{(L)} = {\cal M} \Psi^{(R)} \; .
\label{akc}
\end{equation}
It is given by the ordered product,
\begin{equation}
{\cal M} = 
{\cal D}_{1}{\cal U}_{1,2}{\cal D}_{2}\ldots {\cal U}_{N-1,N}{\cal D}_{N}
\; ,
\label{bkc}
\end{equation}
 of the matrices ${\cal
D}_{k}$ Eq.~(\ref{Mknb}) corresponding to the $k-$th interface,
$k=1,\ldots N$,  and the
evolution matrices ${\cal U}_{k, k+1}$ accounting for the
propagation from  the $k+1$-th to $k$-th potential barrier.

The elements of the S-matrix can be found via the transfer matrix. For
this, we take advantage of the completeness relation in
Eq.~(\ref{Mnnb}) and present ${\cal M}$ as
\begin{equation}
{\cal M}= \sum\limits_{\bm{\mu}, \bm{\mu}'} 
\Psi_{\bm{\mu}}^{(L)}
{\sf M}_{\bm{\mu}\bm{\mu}'}
\Psi_{\bm{\mu}'}^{\ddagger(R)}
\label{fkc}
\end{equation}
where  we denote 
$\bm{\mu}$ the set $(\nu,\sigma )$
and
$\bm{\mu}'= (\nu',\sigma')$. The elements of  the matrix ${\sf M}$ read
\begin{equation}
{\sf M}_{\bm{\mu}\bm{\mu}'} = 
\Psi_{\bm{\mu}}^{\ddagger(L)}
{\cal M}
\Psi_{\bm{\mu}'}^{(R)}
\label{gkc}
\end{equation}
where again $\bm{\mu}$ stands for $(\nu,\sigma)$ and $\Psi_{\nu\sigma
}$ and $\Psi_{\nu,\sigma }$ are the 4-component amplitudes 
Eq.~(\ref{ujc}) and Eq.~(\ref{Mmnb}), respectively.  
The meaning of 
the ${\sf M}$-matrix  is that it is the transfer matrix 
in the plane wave representation:  $C_{\bm{\mu}}^{(L)} =
\sum\limits_{\bm{\mu}'} {\sf
M}_{\bm{\mu}\bm{\mu}'}C_{\bm{\mu}}^{(R)}$ where $C$'s  are the
coefficients in the expansion $\Psi^{(L,R)}=  
\sum\limits_{\bm{\mu}}C_{\bm{\mu}}^{(L,R)}\Psi_{\bm{\mu}}^{(L,R)}$.

Presenting ${\sf M}$, found from Eqs.~(\ref{bkc}),  and (\ref{gkc}), 
in a block form \cite{order},
\begin{equation}
{\sf M}= \left(
\begin{array}{lr}
A   & D   \\
B   & C
\end{array}
\right) \; ,
\label{hkc}
\end{equation}
the $S$-matrix expressed via $2\times 2$ matrices
 $A,B,C,$ and $D$, reads
\begin{equation}
S = 
\left(
\begin{array}{cc}
 B \; A^{-1}& \hspace*{2ex} C -  B A^{-1}  D   \\
A^{-1}&- A^{-1} D
\end{array}
\right)
\label{ikc}
\end{equation}
The matrix element $S_{kn}$ gives the amplitude of the scattering from
the $n-$th incoming state listed in Eq.~(\ref{ckc}) to the $k$-th
final state in Eq.~(\ref{dkc}).

\section{Loop contribution}\label{ss-matrix}

To find contribution of the loops, we first analyze the elementary
process that is the scattering on an isolated barrier.
Consider a single barrier in between two semi-infinite homogeniuos
superconductors left (L), and right (R). The barrier is characterized
by the reflection $r$ and transmission amplitudes $t$. As required by
unitarity, $r t^{*}+ r^{*}t =0$ and $R +T =1$ where $R = |r|^2, T =
|t|^2$. The free wave functions are listed in Eqs.~(\ref{ckc}), 
 and (\ref{dkc}).

The scattering matrix calculated by the method described in in
Appendix \ref{trans}, reads
\begin{equation}
{\sf S}=
\gamma 
\left(
\begin{array}{cccc}
r\beta \beta^{*} & |t|^{2} \alpha^{*}& 
t \beta^{*} & r^{*} t \alpha^{*} \beta \\
 |t|^{2}  \alpha & r^{*}\beta \beta^{*} &
t^{*}r\alpha  \beta^{*} & t^{*}\beta  \\
t \beta & t^{*}r \alpha^{*} \beta  &r \beta \beta^{*}&
|t|^2\alpha \\
r^{*}t \alpha \beta^{*} &  t^{*} \beta^{*} &
\alpha^{*}|t|^{2}&
r^{*}\beta \beta^{*} 
\end{array}
\right).
\label{y81HH}
\end{equation}
where
\begin{equation}
\alpha = - \frac
{
\psi_{-}^{(R)\ddagger} \psi_{+}^{(L)}
}
{
\psi_{-}^{(R)\ddagger} \psi_{-}^{(L)}
}
\;,\;
\beta = - \frac{1}
{
\psi_{+}^{(L)\ddagger} \psi_{+}^{(R)}
} \; , \; 
\gamma = {1\over{1- |r|^{2} \alpha  \alpha^{*} }}
\label{tmc}
\end{equation}
Electron ($\psi_{+}^{(R,L)}$) and hole ($\psi_{-}^{(R,L)}$) wave functions are defined
in Eq.~(\ref{njc}). By their physical meaning,  $\alpha $ is the
amplitude of the Andreev reflection on the SS' interface in the
absence of barrier, and
$\beta $ is the transmission amplitudes. 
Outside the energy gap, 
$
\alpha \beta^{*} + \alpha^{*}\beta =0
\;,\;
\alpha \alpha^{*} + \beta \beta^{*} =1 $  
as required by the quasiparticle current conservation,
As no surprise, the S-matrix in Eq.~(\ref{y81HH}) has the same
structure as that derived in Ref.\onlinecite{She80b} for the NIS interface.

The loop contribution to the Andreev reflection
$A_{e}^{\text{loops}}$ is the sum $A_{e}^{\text{loops}}=
A_{e}^{\text{(a)}} + A_{e}^{\text{(b)}}$ of the processes shown in
Fig.\ref{loops}(a) and (b).  The amplitudes of each of the processes
is the product of factors accumulated along the path. The rules to
find the factors are as follows:

(i) The factor which corresponds to the ballistic part of the
trajectory is $\exp[i \nu p_{\nu }|x_{f}-x_{i}|]$ where $p_{\nu}$ is
the $x-$component of the momentum Eq.~(\ref{4}), $\nu= \pm $ is the
type of the excitation (electron, ``$+$'', or hole, ``$-$'' ), and
$x_{i}$ and $x_{f}$ is the initial and final value of the
$x$-coordinate. One can prove this formula taking into consideration
that the electron propagates in the direction of momentum, and the
hole in the opposite direction.  (The phase accumulated due to a
displacement in the $\bm{p}_{||}$-direction can be omitted since
$\bm{p}_{||}$ is the same for all the ballistic pieces and the path is
closed.)

(ii) For an interface scattering event, the factor is the element of
the S-matrix corresponding to the initial and final states in
Eq.~(\ref{y81HH}).

Looking at Fig. \ref{loops}(a,b) and using these rules, one gets
\begin{eqnarray}
A_{e}^{\text{(a)}} &=& 
S_{23}^{(1)}
e^{i p_{+}a_{1}}
S_{14}^{(2)}
e^{-i p_{-}a_{2}}
S_{22}^{(3)}
e^{-i p_{-}a_{2}}
S_{41}^{(2)}
e^{i p_{+}a_{1}}
S_{31}^{(1)}
\nonumber \\
A_{e}^{\text{(b)}}& =& 
S_{24}^{(1)}
e^{-i p_{-}a_{1}}
S_{23}^{(2)}
e^{i p_{+}a_{2}}
S_{11}^{(3)}
e^{i p_{+}a_{2}}
S_{32}^{(2)}
e^{-i p_{-}a_{1}}
S_{41}^{(1)}
\nonumber 
\end{eqnarray}
where superscript in $S^{(k)}$, $k=1,2,3$, labels the interface (see
Fig.\ref{loops}), and $a_{1}$ ($a_{2}$) is the distance from the
barrier 1 to barrier 2 (from 2 to 3); the momentum $p_{\pm}$ is
calculated for the parameters of the corresponding layer.

Finally, substituting the elements of the S-matrix Eq.~(\ref{y81HH}), 
one gets, 
\begin{eqnarray}
A_{e}^{\text{loops}}     & =  &           
 A_{e}^{\text{(a)}} + A_{e}^{\text{(b)}} 
   \label{vmc}\\   
A_{e}^{\text{loops}}     & =  & {\cal A}\;
\Re \left( 
r_{1} r_{3}^{*} e^{2i \left(\cos \theta p_{F}(a_{1} -a_{2})\right)}
\right)
\label{vmc2}   
\end{eqnarray}
where the coefficient ${\cal A}$ reads
\begin{equation}
  {\cal A}    = 
 - \alpha_{1}
\gamma_{1}^2
\gamma_{2}^2
\gamma_{3}
|
r_{2}t_{1}t_{2}
\alpha_{2}\beta_{1}\beta_{2}\beta_{3}
|^2 
 e^{2i \frac{\cos \theta  }{v_{F}}(\xi_{1}a_{1} + \xi_{2}a_{2})} \;.
              \label{vmc3}
\end{equation}
Here, $\theta $ is the angle between the direction of the trajectory
and the $x-$axis, and $\xi_{1,2}$ is defined by Eq.~(\ref{mjc}) for
the corresponding layer.


\begin{thebibliography}{99}
\bibitem{neglect} In Eq.~(\ref{2kc}), we do not show the factor
$(p_{+}+ p_{-})/ \sqrt{4 p_{+} p_{-}}$ which equals to 1 in the main
approximation with respect to the small parameter $\Delta /E_{F}$,
$E_{F}$ being the Fermi energy.
\bibitem{generalD} One can show that the most general form of ${\cal
D}$ compatible with the current conservation and the time reversal
symmetry is ${\cal D}= \left(
\begin{array}{lr}
a   & id   \\
 ib  & c
\end{array}
\right)$ 
where $a,b,c$, and $d$ are real parameters satisfying the condition
$\det {\cal D}=1$.
\bibitem{order} 
The basis states $\Psi_{\bm{\mu}}^{(L,R)}$ are ordered as follows.
On the left and right, 
1: $\Psi_{++}$,
2: $\Psi_{--}$,
3: $\Psi_{+-}$,
4: $\Psi_{-+}$.
\end{thebibliography}
\end{document}